\documentclass[letterpaper,journal]{IEEEtran}
\usepackage{amsmath,amsfonts}
\allowdisplaybreaks
\usepackage{algorithm}
\usepackage{algpseudocode}
\algrenewcommand\algorithmicrequire{\textbf{Input:}}
\algrenewcommand\algorithmicensure{\textbf{Output:}}
\usepackage{array}
\usepackage[caption=false,font=normalsize,labelfont=sf,textfont=sf]{subfig}
\usepackage{textcomp}
\usepackage{stfloats}
\usepackage{url}
\usepackage{verbatim}
\usepackage{graphicx}
\usepackage{cite}
\usepackage{xcolor}
\usepackage{amssymb}
\usepackage{cuted}
\usepackage{enumitem}
\usepackage{float}
\usepackage{booktabs, multirow}
\usepackage{tabularx}
\usepackage{amssymb,amsmath}
\usepackage{xspace}
\usepackage{bm}
\usepackage{bbm}
\usepackage{mathtools}

\let\originalleft\left
\let\originalright\right
\renewcommand{\left}{\mathopen{}\mathclose\bgroup\originalleft}
\renewcommand{\right}{\aftergroup\egroup\originalright}

\newcommand{\mat}[1]{\mathbf{#1}} 
\renewcommand{\vec}[1]{\mathbf{#1}} 
\newcommand{\vecfd}[1]{\bm{\mathsf{#1}}} 

\newcommand{\parens}[1]{{\left(#1\right)}\xspace}

\newcommand{\complex}{{\mathbb{C}}\xspace}

\newcommand{\ind}[1]{\mathbbm{1}\parens{#1}}

\def\va{{\vec{a}}}

\def\vw{{\vec{w}}}
\def\vx{{\vec{x}}}
\def\vy{{\vec{y}}}

\def\vfu{{\vecfd{u}}}
\def\vfv{{\vecfd{v}}}
\def\vfw{{\vecfd{w}}}
\def\vfx{{\vecfd{x}}}
\def\vfy{{\vecfd{y}}}

\def\mH{{\mat{H}}}

\begin{document}

\title{Timing-Aware Satellite Association \\ for Multi-LEO Direct-to-Handset Communications}

\author{Hyunwoo Lee, Incheol Hwang,~\IEEEmembership{Member,~IEEE,} and Daesik Hong,~\IEEEmembership{Fellow,~IEEE}
\thanks{This work was supported in part by the National Research Foundation of Korea (NRF) through the Korea government (MSIT) under Grant 2022R1A2C300415413.

H.~Lee is with Samsung Electronics, Suwon 16677, South Korea (e-mail: gksdnrh27@yonsei.ac.kr).

I.~Hwang is with the Agency for Defense Development, Daejeon 34060, South Korea.

D.~Hong is with the Information Telecommunication Lab (ITL), School of Electrical and Electronic Engineering, Yonsei University, Seoul, South Korea (e-mail: daesikh@yonsei.ac.kr).}
}



\maketitle

\begin{abstract}
The rapid deployment of large-scale low Earth orbit (LEO) satellite constellations has positioned direct-to-handset (D2H) communications as a key enabler of future non-terrestrial networks.
However, the limited link budget of handheld devices makes broadband service delivery challenging, and multi-satellite cooperative transmission is often required to provide sufficient power gain.
In practice, such cooperation is severely hindered by asynchronous reception across satellites.
This paper analyzes the received-signal model under the 3rd Generation Partnership Project (3GPP) transmitter structure and shows that satellite-dependent propagation delays prevent simultaneous timing alignment for multiple user terminals (UTs).
This timing mismatch induces severe inter-carrier interference (ICI) and inter-symbol interference (ISI), even from the intended signals, which fundamentally constrains the achievable cooperative gain.
To address this issue, we propose a timing-aware satellite association strategy that enables cooperation only with satellites expected to satisfy a UT-side timing tolerance, thereby avoiding dominant asynchronous interference by design.
Simulation results demonstrate that the proposed strategy improves throughput performance compared to single-satellite transmission and fully connected multi-satellite baselines.
\end{abstract}

\begin{IEEEkeywords}
6G, satellite communication (SatCom), low Earth orbit (LEO), non-terrestrial network (NTN), direct-to-handset (D2H), cooperation, association.
\end{IEEEkeywords}

\section{Introduction}\label{sec_1}
Dense low Earth orbit (LEO) constellations have emerged as a key enabler of next-generation communication systems, as they can provide near-global coverage while ensuring relatively high communication quality compared to other non-terrestrial network (NTN) platforms. 
Recently, \textit{direct satellite-to-handset (D2H) communication} has gained increasing attention, where LEO satellites directly connect to smartphones to deliver data services. 
Unlike conventional satellite systems that require dedicated terminals, D2H communication allows users to access data services anywhere on Earth with their existing smartphones~\cite{WCM23_D2C, COMMAG25_D2D}. 
At present, its use cases are limited to voice communication~\cite{WESCON95_IRIDIUM}, short messaging service (SMS), and SOS alerts~\cite{SMS_service}. 
However, in the near future, broadband Internet services---traditionally offered via terrestrial networks---are also expected to be delivered through LEO satellites, paving the way toward a unified space--terrestrial network~\cite{OJCOMS23_D2D}.

In D2H communication, the user terminal (UT) is a handset device, which inherently limits the link budget and makes it difficult to ensure sufficient received power~\cite{WCM23_D2C}. 
While services such as voice communication, SMS, and SOS alerts do not demand high data rates and can therefore be supported without major issues, the situation differs for broadband Internet services. 
A single satellite alone cannot provide the required data rate to handset devices, making multi-satellite cooperative systems with joint transmission or multiplexing indispensable~\cite{OJCOMS23_D2D}.
Meanwhile, with the deployment of large-scale LEO constellations by various satellite operators, including SpaceX's Starlink, Amazon's Kuiper Project, and OneWeb, the feasibility of multi-satellite cooperative systems has been further enhanced~\cite{COMST23_LEOMIMO}.

However, despite this potential, significant challenges arise in practice, particularly due to propagation delay differences across satellites. 
In LEO satellite communication (SatCom) systems, the long communication distance leads to substantial propagation delay, and signals transmitted from different satellites to the same UT experience distinct delays with large differences~\cite{COMMAG26_LEOTDD}. 
For a single-user scenario, this issue can be mitigated through satellite synchronization. 
Yet, when multiple users are served simultaneously, the problem becomes much more severe. 
Considering the multiple input and multiple output (MIMO)-orthogonal frequency division multiplexing (OFDM) transmitter architecture specified in the 3rd Generation Partnership Project (3GPP) standards~\cite{dahlman2020nr,3gpp38211}, the aggregate signal transmitted by the satellites is temporally synchronized across all users. 
In other words, when the transmission is aligned to enable one user to receive a perfectly synchronized signal, other users inevitably suffer from severe asynchronous reception. 
Such misalignment, although originating from desired signals, induces inter-carrier interference (ICI) and inter-symbol interference (ISI), resulting in significant performance degradation. 
Consequently, although joint transmission is designed to overcome the low link budget, its achievable power gain is fundamentally constrained by severe ISI and ICI resulting from asynchronous reception.
This observation highlights the need for schemes that can mitigate the severe performance degradation caused by asynchronous reception while achieving sufficient power gain.

To mitigate the performance loss caused by asynchronous reception while retaining cooperative power gain, two approaches are possible.
The first is to avoid dominant asynchronous interference by design, i.e., to cooperate only with satellites whose delay offsets can be aligned within a tolerable UT-side timing margin.
The second is to suppress the resulting interference through interference management techniques, such as cancellation and receiver-side compensation.
In this paper, we adopt the first approach and propose a timing-aware satellite association strategy that selects only such satellites, enabling multi-LEO cooperation without additional timing synchronization among satellites while still achieving substantial cooperative power gain.

The rest of this paper is organized as follows.
Section~\ref{sec_2} reviews related work and clarifies our contributions.
After introducing the multi-satellite multi-UT system model in Section~\ref{sec_3}, Section~\ref{sec_4} formulates the received signal under asynchronous reception to quantify the resulting ICI and ISI.
Section~\ref{sec_5} then presents the proposed timing-aware satellite association strategy.
Simulation results and discussions are provided in Section~\ref{sec_6}, followed by concluding remarks in Section~\ref{sec_7}.

\section{Related Work and Contributions}\label{sec_2}
\subsection{Related Work}
With the explosive growth of the space industry, research interest in LEO SatCom has significantly increased, and numerous studies have been conducted in this area. 
To propose satellite association strategies for multi-LEO D2H communications, we first summarize the current state of research on transmission techniques in LEO SatCom systems.

\subsubsection{\textbf{Single-Satellite Communication Systems}}
MIMO has become one of the core communication techniques in cellular systems, as it provides multiple benefits such as power gain, multiplexing gain, and diversity gain. 
Motivated by these advantages on the ground, many studies have attempted to apply MIMO techniques to SatCom systems in order to achieve similar performance improvements. 
To capture the line-of-sight (LoS)-dominant characteristics of LEO SatCom systems, the authors of \cite{JSAC20_singleSAT} presented a channel model tailored to such environments.
Using this model, they applied a precoder that maximizes the signal-to-leakage-plus-noise ratio (SLNR) and further demonstrated that statistical channel state information (CSI) can be sufficient in SatCom environments.
The work in \cite{TCOM22_LOS} focused on transmit beamformer design and proved that the rank-one property of LEO SatCom channels makes single-stream transmission to each UT the optimal choice.
Moreover, in \cite{TB23_singleSAT}, an angle-based beamforming method was proposed by leveraging ephemeris and global navigation satellite system (GNSS) information. 
These works established the foundation of MIMO techniques in LEO SatCom systems by adapting terrestrial MIMO methods to satellite-specific environments. 
However, their scope remains limited to single-satellite scenarios.

Beyond the conventional digital and hybrid beamforming, holographic MIMO has recently emerged as a practical architecture that realizes a large aperture with densely packed low-cost metamaterial elements \cite{WCM20_HMIMO}.
Two representative implementations are the stacked intelligent metasurface (SIM), which performs beamforming in the wave domain as the signal propagates through multiple cascaded metasurface layers \cite{COMMAG26_SphericalSIM, TCOM25_SIM_NearField}, and the reconfigurable holographic surface (RHS), which generates directional beams based on the holographic principle \cite{WCM21_RHS, TWC24_RHS_CellFree}.
The SIM has also been mounted on an LEO satellite to perform multi-user beamforming in the wave domain relying on statistical CSI \cite{WCL24_SIM_LEO}.
Note that these works focus on the implementation of the beamformer at each node, whereas the ICI and ISI considered in this paper originate from the propagation delay differences among satellites.
Consequently, the proposed timing-aware association can be applied irrespective of the beamforming architecture.

\subsubsection{\textbf{Multi-Satellite Communication Systems in Narrowband Scenarios}}
Since a ground location is typically covered by two or more satellites, the prerequisites for multi-satellite connectivity are already satisfied~\cite{COMST23_LEOMIMO}.
Accordingly, a variety of studies have been conducted with the aim of realizing multi-satellite communication systems.

In the early stages, research primarily focused on defining and analyzing the performance gains achievable through multi-satellite cooperation. 
A cooperative beamforming scenario involving geostationary orbit (GEO) satellites was presented in \cite{Globecom14_MultiGEO}, and the resulting cooperative transmission gains and performance improvements were analyzed.
A model in which two LEO satellites simultaneously transmit to a multi-antenna terminal was introduced in \cite{Access20_MultiLEO_narrow}, and the throughput and outage probability were examined as functions of the terminal antenna configuration and satellite positions.
For the uplink, a statistical CSI-based cooperative detection model for multi-LEO satellites was developed in \cite{PIMRC23_MultiLEO_narrow}, and its ergodic capacity was evaluated.

Building upon these initial efforts, subsequent studies have investigated more advanced MIMO techniques, particularly in environments where multiple satellites in dense LEO constellations form distributed massive MIMO (DM-MIMO) structures. 
A DM-MIMO architecture, together with power allocation and handover management schemes, was proposed in \cite{OJCOMS22_MultiLEO_narrow}. 
As a follow-up, they demonstrated the potential of DM-MIMO to enhance the link budget of handset devices and analyzed the performance gain using real Starlink data~\cite{OJCOMS23_D2D}. 
In \cite{TWC24_MultiLEO_narrow}, the authors presented a low-complexity transmission framework that combines hybrid beamforming with user scheduling in multi-satellite cooperative networks. 
More recently, a user-centric cooperative framework that reduces handover frequency 
while improving both spectral efficiency and coverage has been proposed in \cite{JSAC25_MultiLEO_narrow}.

Thus, a wide range of studies have investigated multi-satellite cooperative systems, with the majority focusing on applying MIMO techniques to improve throughput. 
However, these works remain at an early stage, as they either neglect the practical issue of asynchronous reception or simplify the system by assuming narrowband models.

\subsubsection{\textbf{Multi-Satellite Communication Systems in Wideband Scenarios}}
While narrowband studies have primarily focused on throughput improvements under simplified assumptions, more recent works on wideband systems have begun to highlight practical challenges, particularly those arising from asynchronous reception.
In \cite{TWC24_MultiLEO_wide}, the authors highlighted the problem of multi-user interference (MUI) caused by asynchronous reception due to propagation delay differences between satellites and UTs. 
To mitigate this interference, they proposed a cooperative beamforming architecture in a single-carrier setting based on the weighted minimum mean square error (WMMSE) criterion. 
In \cite{TVT25_MultiLEO_wide}, the authors considered distributed beamforming in a multi-LEO SatCom environment employing OFDM. 
In addition to asynchronous interference resulting from propagation delay differences, they also examined the impact of imperfect delay and Doppler shift compensation, and addressed it using an SLNR-maximizing precoder. 

These works explicitly recognize the asynchronous reception problem inherent in realistic SatCom environments.
However, they either assume single-carrier transmission for wideband LEO links or rely on perfectly controlled transmit timing so that each receiver obtains its desired signal with perfectly aligned timing.
Consequently, the loss of subcarrier orthogonality in OFDM systems and the practical MIMO-OFDM transceiver architecture specified in 3GPP standards are not captured.

\subsection{Contributions}
Existing literature on multi-LEO satellite cooperative systems has proposed various beamforming techniques, but most studies remain at an early stage.
They largely overlook practical asynchronous interference and instead focus on the theoretical performance gains achievable with multi-satellite cooperation.
More recent works have considered wideband systems and asynchronous interference, but they do not account for the transceiver architecture specified in 3GPP standards, nor do they capture the loss of OFDM subcarrier orthogonality caused by asynchronous reception.

This paper investigates interference signals induced by asynchronous reception in practical multi-LEO cooperative communication systems that incorporate standard-compliant transceiver architectures. 
We also propose satellite association strategies to mitigate such interference, thereby addressing the low link budget problem inherent in D2H communications. 
The specific contributions of this paper can be summarized as follows.

\begin{itemize}
    \item We propose a satellite association strategy and frame structure for multi-LEO D2H communications. 
    Specifically, we exploit an additional cyclic prefix (CP) margin dedicated to multi-satellite connectivity and attempt to associate only with satellites whose signals arrive within this CP window. 
    Joint transmission from the associated satellites then becomes effectively free from asynchronous-reception-induced interference, allowing the receiver to fully exploit the power gain from multiple satellites.
    \item We develop a mathematical analysis framework for multi-LEO satellite cooperative communication systems operating under a 3GPP-compliant MIMO-OFDM transmitter architecture. 
    In particular, we establish a discrete-time signal model for the system throughput that explicitly accounts for ICI and ISI caused by asynchronous reception. 
    Based on this model, we derive the average throughput of the proposed satellite association strategy, and we validate the accuracy of the analysis through simulation results.
    \item We propose a downlink synchronization algorithm tailored to the proposed satellite association strategy in order to maximize throughput. 
    The algorithm adjusts the downlink timing reference at the UT so that the received signals from as many satellites as possible fall within the CP window. 
    This enables the UT to aggregate power from a larger set of satellites and thereby achieve higher throughput.
    \item We present simulation results under a 3GPP-specified new radio (NR) NTN parameter setting to evaluate the proposed satellite association strategy.
    For benchmarking, two baselines are considered: (i) single-satellite transmission and (ii) a fully connected scheme where each user is served by all visible satellites without association control.
    The proposed strategy achieves higher throughput than both baselines, demonstrating substantial cooperative power gain while avoiding dominant asynchronous inter-satellite interference.
\end{itemize}

\section{System Model}\label{sec_3}
\begin{figure}[!t]
    \centering
    \includegraphics[width=\columnwidth]{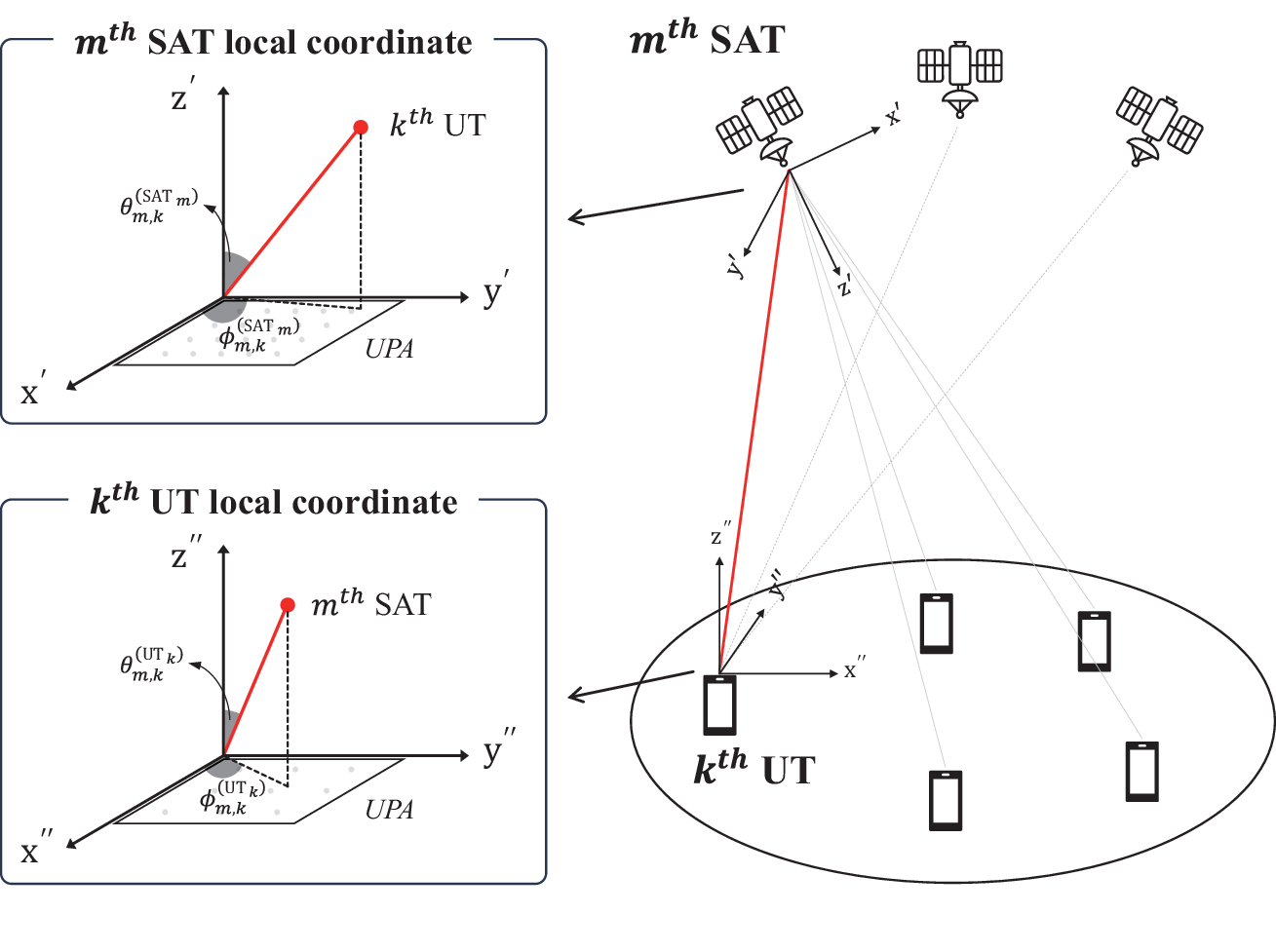}
    \caption{System model of multi-LEO satellite D2H communication, consisting of $M$ LEO satellites and $K$ UTs. Each satellite and UT is equipped with a UPA.}
    \label{fig:system_model}
\end{figure}

We consider a system consisting of $M$ LEO satellites and $K$ UTs, as illustrated in Fig.~\ref{fig:system_model}. 
Each satellite is equipped with a uniform planar array (UPA) composed of $N_{\mathrm{tx},\mathrm{x}}$ elements along the x-axis and $N_{\mathrm{tx},\mathrm{y}}$ elements along the y-axis, yielding a total of $N_{\mathrm{tx}} = N_{\mathrm{tx},\mathrm{x}} \times N_{\mathrm{tx},\mathrm{y}}$ elements. 
Similarly, each UT is equipped with a UPA consisting of $N_{\mathrm{rx}} = N_{\mathrm{rx},\mathrm{x}} \times N_{\mathrm{rx},\mathrm{y}}$ elements. 
The UPA normal of each satellite points toward the Earth's center, while the UPA normal of each UT points outward from the Earth's center. 
In their respective local coordinate systems, the UPAs are placed on the xy-plane with the z-axis representing the normal vector.

As specified in 3GPP Release 17 (Rel-17), we assume that the Doppler shift can be perfectly compensated using the satellite's GNSS functionality~\cite{38.821}.
Under the LoS-dominant NTN channel, the root-mean-square delay spread is a few to tens of nanoseconds and the Rician $K$-factor is about 10~dB~\cite{38.811}, so the channel variation is dominated by the deterministic motion of the satellites, which can be predicted from the ephemeris.
Considering a total of $N$ subcarriers, the channel corresponding to the $n$-th subcarrier between the $m$-th satellite and the $k$-th UT can be expressed as in~\cite{TB23_3Dchannel, TWC25_WideSatCom}:
\begin{equation}
\begin{aligned}
\mH_{m,k}^{n}
&= \sqrt{\beta_{m,k}}
   \sum_{l=1}^{L_\mathrm{p}} \alpha_{m,k,l}^n\,
   \va^{\mathrm{UT},n}_{l}
   \bigl(\theta^{(\mathrm{UT}_k)}_{m,k,l},
         \phi^{(\mathrm{UT}_k)}_{m,k,l}\bigr) \\
&\quad\times
   \left\{
   \va^{\mathrm{SAT},n}_{l}
   \bigl(\theta^{(\mathrm{SAT}_m)}_{m,k,l},
         \phi^{(\mathrm{SAT}_m)}_{m,k,l}\bigr)
   \right\}^\mathrm{H},
\end{aligned}
\label{eq:original_channel}
\end{equation}
where $\mH_{m,k}^{n}\in\complex^{N_{\mathrm{rx}}\times N_{\mathrm{tx}}}$, $\beta_{m,k}$ denotes the path loss between the $m$-th satellite and the $k$-th UT, and $L_{\mathrm{p}}$ represents the number of propagation paths.
$\alpha_{m,k,l}^n$ is the complex gain of the $l$-th propagation path between the $m$-th satellite and the $k$-th UT.
$\va^{\mathrm{UT}}_{l}$ and  $\va^{\mathrm{SAT}}_{l}$ are the array response vectors of the UT and the satellite, respectively, associated with the $l$-th propagation path.
The array response vector for an object $o \in \{\mathrm{SAT}, \mathrm{UT}\}$ can be expressed as
\begin{equation}
    \va_l^{o,n}\left(\theta^{(o)}_{m,k,l}, \phi^{(o)}_{m,k,l}\right)=\va^n\left(\vartheta^{o,\mathrm{x}}_{m,k,l}\right) \otimes\va^n\left(\vartheta^{o,\mathrm{y}}_{m,k,l}\right),\label{eq:array_response_vector1}
\end{equation}
where $\vartheta_{m,k,l}^{o,\mathrm{x}} = \sin{\theta^{(o)}_{m,k,l}} \cos{\phi^{(o)}_{m,k,l}}$ and $\vartheta_{m,k,l}^{o,\mathrm{y}} = \sin{\theta^{(o)}_{m,k,l}} \sin{\phi^{(o)}_{m,k,l}}$, $\theta$ and $\phi$ represent the zenith and azimuth angles, respectively, in the local coordinate system, and $\otimes$ denotes the Kronecker product.
The array response vector along the $d \in \{\mathrm{x,y}\}$ axis can be expressed as
\begin{equation}
\begin{aligned}
\va^n\bigl(\vartheta^{o,d}_{m,k,l}\bigr)
&= \frac{1}{\sqrt{N_{o,d}}}
\Bigl[1,\;
e^{-j2\pi \frac{f_n}{c} d_{\mathrm{s}}\vartheta^{o,d}_{m,k,l}},\\
&\qquad\ldots,\;
e^{-j2\pi \frac{f_n}{c} d_{\mathrm{s}}(N_{o,d}-1)\vartheta^{o,d}_{m,k,l}}
\Bigr],
\end{aligned}
\label{eq:array_response_vector2}
\end{equation}
where $f_n$ denotes the frequency of the $n$-th subcarrier and $c$ is the speed of light. 
$d_{\mathrm{s}}$ represents the antenna spacing of the UPA, which is typically set to half of the wavelength. 
Since scattering objects are scarce in LEO SatCom systems, the array response vectors corresponding to different propagation paths between the $m$-th satellite and the $k$-th UT are nearly identical~\cite{TCOM22_LOS}. 
Therefore, we can set $L_\mathrm{p}=1$, and as a result, \eqref{eq:original_channel} can be expressed in a simplified form as follows.
\begin{equation}
\begin{aligned}
\mH_{m,k}^{n}
&= \sqrt{\beta_{m,k}} \,\alpha_{m,k}^n
   \va^{\mathrm{UT},n}\!\bigl(\theta^{(\mathrm{UT}_k)}_{m,k},
                               \phi^{(\mathrm{UT}_k)}_{m,k}\bigr) \\
&\quad\times
   \left\{\va^{\mathrm{SAT},n}\!\bigl(\theta^{(\mathrm{SAT}_m)}_{m,k},
                                      \phi^{(\mathrm{SAT}_m)}_{m,k}\bigr)\right\}^\mathrm{H} .
\end{aligned}
\label{eq:channel}
\end{equation}
Once a link between the satellites and the UT is established, data service is provided through the channel defined in \eqref{eq:channel}. 
In the following section, we model the transmit and receive signals based on the defined system model and examine the impact of asynchronous reception.

\section{Signal Model under Asynchronous Reception}\label{sec_4}
\begin{figure}[!t]
    \centering
    \includegraphics[width=1\columnwidth]{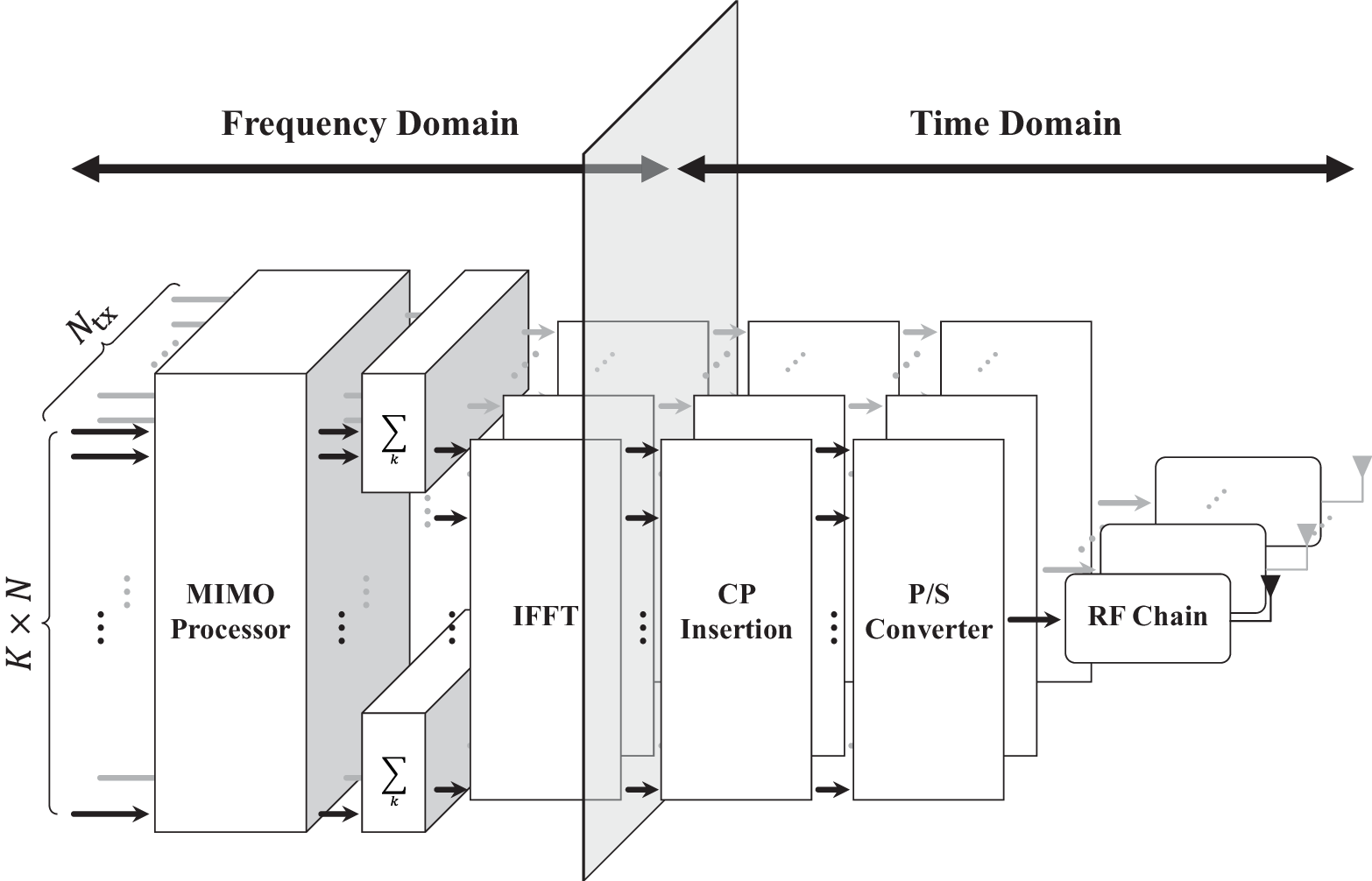}
    \caption{MIMO-OFDM transmitter architecture as specified in 3GPP NR~\cite{dahlman2020nr, 3gpp38211}.}
    \label{fig:3GPP_transmitter}
\end{figure}

\begin{figure}[!t]
\centering
\includegraphics[width=1\linewidth]{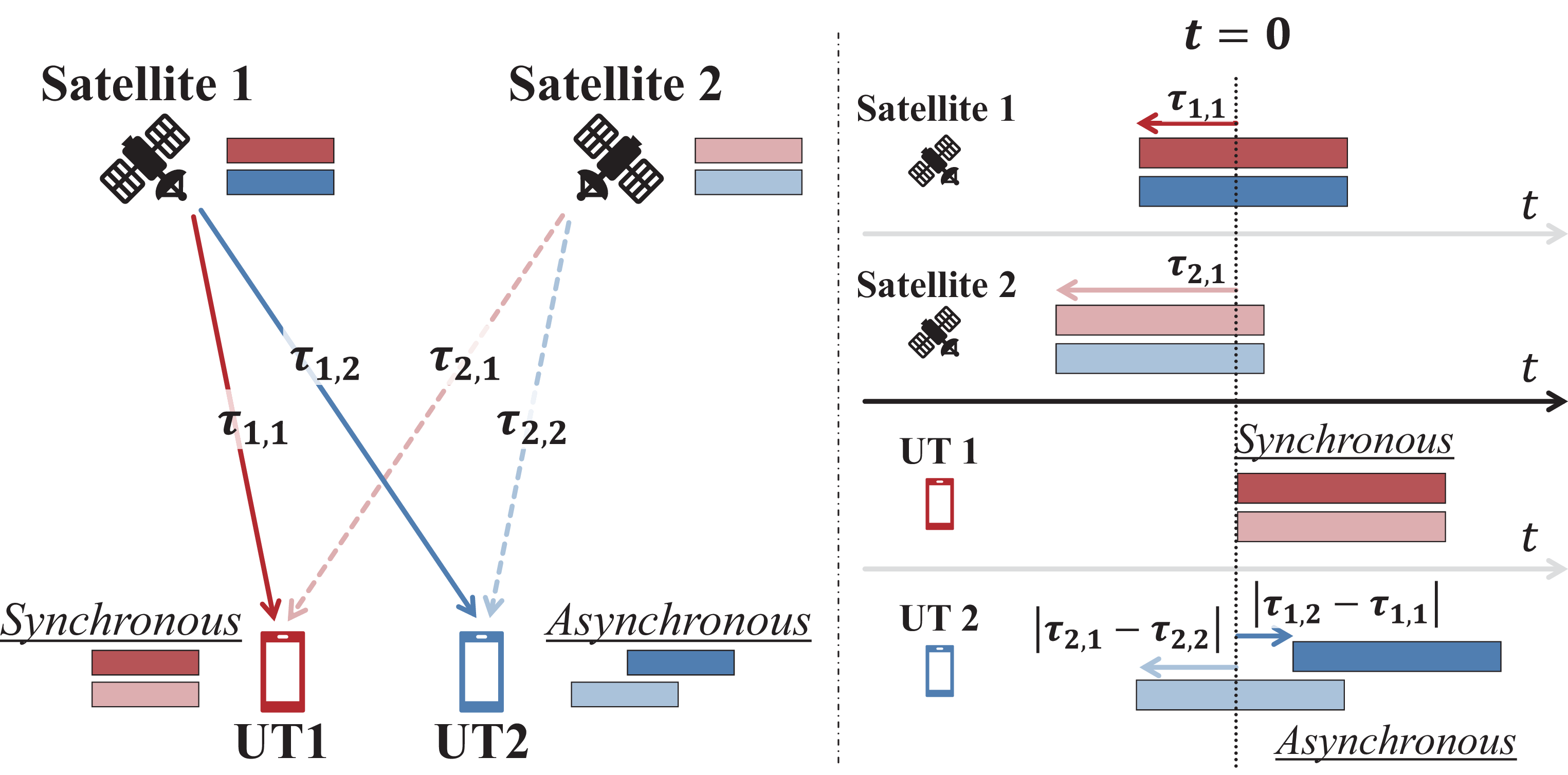}
\caption{Illustration of downlink timing adjustment under the MIMO-OFDM architecture with two satellites and two UTs.}
\label{fig:asynchronous_example}
\end{figure}

The MIMO-OFDM transmitter architecture specified by 3GPP is shown in Fig.~\ref{fig:3GPP_transmitter}~\cite{dahlman2020nr,3gpp38211}.
This architecture was introduced in long term evolution (LTE) and has been retained in 5G NR.
In this architecture, MIMO precoding is performed in the frequency domain, and OFDM modulation---comprising the inverse fast Fourier transform (IFFT), CP insertion, and parallel-to-serial conversion---follows, after which the radio-frequency (RF) chains generate the per-antenna waveforms.
A key implication of this architecture is that user-specific timing offsets cannot be applied at the transmitter, since precoding occurs in the frequency domain and all UT signals are emitted with a common timing reference.
Consequently, in cooperative transmission, if the cooperating satellites adjust their timing so that one UT receives perfectly aligned desired signals, the other UTs inevitably experience asynchronous reception~\cite{TCOM19_Async}, as illustrated in Fig.~\ref{fig:asynchronous_example}.
It should be noted that the timing advance scheme introduced for NTN in 3GPP Rel-17 and enhanced in Rel-18~\cite{3gpp38213} is a UT-side pre-compensation for the uplink based on the GNSS position of the UT and the satellite ephemeris, and does not alter the downlink transmission.
Although each satellite may adjust its downlink transmission timing, the adjustment is applied to the common time-domain waveform and hence to all UTs, so that the asynchronous reception described above remains under the Rel-17/18 scheme.

In terrestrial cellular systems, the propagation delay between a base station (BS) and a UT is relatively small, and the variation across different links is also limited. 
As a result, in cooperative transmission scenarios, the resulting time misalignment typically remains within the CP of an OFDM symbol. 
Such misalignment does not break OFDM subcarrier orthogonality and can be effectively handled by simple phase alignment. 
In contrast, in SatCom systems, the propagation delay between a satellite and a UT is significantly larger, and the variations across different links are substantial. 
Consequently, the time misalignment often exceeds the CP duration, leading to severe ISI and ICI. 
These observations motivate the following signal and interference modeling for multi-LEO cooperative transmission under a 3GPP-compliant MIMO-OFDM transmitter architecture.

\subsection{Transmitted/Received Signal}
We consider a system in which $M$ satellites simultaneously serve $K$ UTs through joint transmission. 
For a given UT, different satellites transmit the same stream, and all UTs share the entire frequency resource, with different streams assigned to different subcarriers. 
Due to the characteristics of SatCom systems, both the absolute propagation delay and the delay differences among satellites exceed the duration of an OFDM symbol. 
However, since OFDM symbol-level synchronization among satellites can be maintained, only sample-level timing offsets are considered. 
Such symbol-level synchronization can be achieved through GNSS-based timing~\cite{38.821} or inter-satellite time synchronization techniques~\cite{OJCOMS22_MultiLEO_narrow}.
In other words, the signals received from different satellites may experience sample-level misalignment within a single OFDM symbol. 
Therefore, all subsequent timing offsets in this paper refer to the sample-level misalignment remaining after symbol-level synchronization.

\begin{figure}[!t]
    \centering
    \includegraphics[width=1\columnwidth]{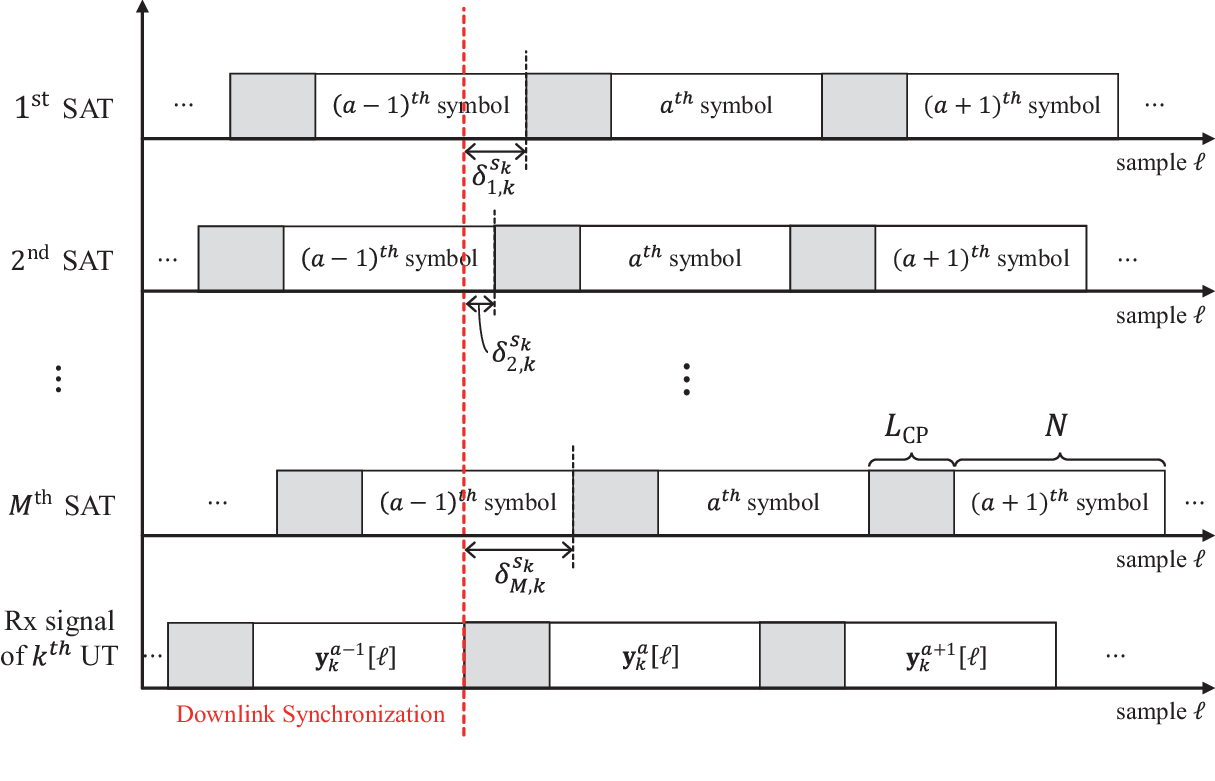}
    \caption{Illustration of asynchronous reception at the $k$-th UT, where propagation delay differences among multiple satellites cause sample-level misalignment of the received signals.}
    \label{fig:asynchronous_reception}
\end{figure}

In the frequency domain, the transmitted signal of the $m$-th satellite on the $n$-th subcarrier is expressed as the superposition of the signals intended for all served UTs, as follows:
\begin{equation}
\vfx_{m}^{a,n} = \sum_{k=1}^{K} \vfv_{m,k}^{n} d_{k}^{a,n}, 
\quad \vfv_{m,k}^{n} \in \complex^{N_{\mathrm{tx}} \times 1}. \label{eq:tx_signal_frequency}
\end{equation}
Here, $d_k^{a,n}$ is the $a$-th data stream of the $k$-th UT on the $n$-th subcarrier and $\vfv_{m,k}^{n}$ denotes the precoding vector of the $m$-th satellite for the $k$-th UT on the $n$-th subcarrier. 
In practice, due to OFDM symbol synchronization, the symbol index may differ across UTs. 
However, for notational simplicity and without loss of generality, we represent the transmission as if all UTs receive the $a$-th OFDM symbol simultaneously. 
The transmitted signal in the time domain can then be expressed as
\begin{equation}
\vx_{m}^{a}[\ell] 
= \frac{1}{\sqrt{N}} \sum_{n=0}^{N-1} \vfx_{m}^{a,n} 
   e^{j 2\pi n \ell/{N}}, \quad \ell = 0,1,\ldots,N-1. \label{eq:tx_signal_time}
\end{equation}
Once the CP of length $L_{\mathrm{CP}}$ is inserted, the signal \eqref{eq:tx_signal_time} is transmitted through the satellite antennas to the UTs. 
Due to propagation delay differences across satellites, the $k$-th UT then receives misaligned signals, as illustrated in Fig.~\ref{fig:asynchronous_reception}.
The sample offset between the $m$-th satellite and the $k$-th UT is denoted by $\delta_{m,k}^{s_k}$, where $s_k \in [0,N-1]$ is the downlink synchronization point. 
It is defined as
\begin{equation}
    \delta_{m,k}^{s_k} = \left(\left\lfloor\frac{\tau_{m,k}}{T_{\mathrm{s}}}\right\rfloor -s_k\right) \bmod \left(N+L_{\mathrm{CP}}\right), \label{eq:sample_offset}
\end{equation}
where $\tau_{m,k}$ denotes the propagation delay and $T_{\mathrm{s}}$ is the sampling period.
Then the time-domain received signal after CP removal can be expressed as
\begin{align}
\vy_{k}^{a}[\ell] 
&= \sum_{m=1}^{M} \mH_{m,k}^n \Big(
      \vx_{m}^{a}\big[(\ell-\delta_{m,k}^{s_k}) \bmod N\big]\,
      \mathbbm{1}_{[\Delta_{m,k}^{s_k},\,N-1]}[\ell] \notag\\
&\qquad
      + \vx_{m}^{a-1}\big[(\ell-\delta_{m,k}^{s_k}+N) \bmod N\big]\,
      \mathbbm{1}_{[0,\,\Delta_{m,k}^{s_k}-1]}[\ell]
   \Big) \notag\\
&\quad+ \vw_{k}^{a}[\ell],
\label{eq:rx_signal_time}
\end{align}
\noindent where $\Delta_{m,k}^{s_k}\triangleq\max(\delta_{m,k}^{s_k}-L_{\mathrm{CP}},0)$ represents the effect of CP insertion and removal, and $\vw_{k}^{a}[\ell]$ is additive white Gaussian noise with zero mean and variance $\sigma_w^2$. The indicator function $\mathbbm{1}_{[\cdot]}[\cdot]$ is defined to capture the contributions from both the previous and current OFDM symbols due to asynchronous reception, as
\begin{equation}
\mathbbm{1}_{[a,b]}[\ell] =
\begin{cases}
1, & a \leq \ell \leq b, \\
0, & \text{otherwise}.
\end{cases} \label{eq:indicator_function}
\end{equation}
After passing through the fast Fourier transform (FFT) block at the receiver, the frequency-domain received signal can be expressed as
\begin{equation}
\vfy_{k}^{a,n} 
= \frac{1}{\sqrt{N}} \sum_{\ell=0}^{N-1} 
\vy_{k}^{a}[\ell] \, e^{-j {2\pi \ell n}/{N}}. \label{eq:rx_signal_frequency_mid}
\end{equation}
For notational convenience and to clearly distinguish the effects of ICI and ISI, 
we define the ICI leakage coefficient $A_{nn'}(\delta)$ and the ISI leakage coefficient $B_{nn'}(\delta)$ as
\begin{equation}
\begin{split}
A_{nn'}(\delta) 
&= \frac{1}{N} e^{-j \tfrac{2\pi n' \delta}{N}}
\sum_{\ell=\Delta}^{N-1} e^{\,j \tfrac{2\pi}{N}(n'-n)\ell} \\
&=
\begin{cases}
\displaystyle \frac{N-\Delta}{N} e^{-j \tfrac{2\pi n' \delta}{N}}, & n = n', \\[10pt]
\displaystyle \frac{1}{N} e^{-j \tfrac{2\pi n' \delta}{N}}
\frac{1 - e^{\,j \tfrac{2\pi}{N}(n'-n)(N-\Delta)}}{1 - e^{\,j \tfrac{2\pi}{N}(n'-n)}}, & n \neq n'.
\end{cases}
\end{split}
\label{eq:ICI_leakage_coefficient}\tag{11}
\end{equation}
\begin{equation}
\begin{split}
B_{nn'}(\delta) 
&= \frac{1}{N} e^{-j \tfrac{2\pi n' \delta}{N}}
\sum_{\ell=0}^{\Delta-1} e^{\,j \tfrac{2\pi}{N}(n'-n)\ell} \\
&=
\begin{cases}
\displaystyle \frac{\Delta}{N} e^{-j \tfrac{2\pi n' \delta}{N}}, & n = n', \\[10pt]
\displaystyle \frac{1}{N} e^{-j \tfrac{2\pi n' \delta}{N}}
\frac{1 - e^{\,j \tfrac{2\pi}{N}(n'-n)\Delta}}{1 - e^{\,j \tfrac{2\pi}{N}(n'-n)}}, & n \neq n'.
\end{cases}
\end{split}
\label{eq:ISI_leakage_coefficient}\tag{12}
\end{equation}
\begin{figure*}[!b]
\centering
\hrule
\begin{align}
\vfy_{k}^{a,n}
&=
\underbrace{\sum_{m=1}^{M} \mathbf{G}_{m,k}^{n,n}\vfv_{m,k}^{n} d_{k}^{a,n}}_{\text{desired signal}}
+\underbrace{\sum_{k'\ne k}\sum_{m=1}^{M} \mathbf{G}_{m,k}^{n,n}\vfv_{m,k'}^{n} d_{k'}^{a,n}}_{\text{MUI}}
+\underbrace{\sum_{n'\ne n}\sum_{m=1}^{M} \mathbf{G}_{m,k}^{n,n'}\vfv_{m,k}^{n'} d_{k}^{a,n'}}_{\text{ICI (self)}}
+\underbrace{\sum_{n'\ne n}\sum_{k'\ne k}\sum_{m=1}^{M} \mathbf{G}_{m,k}^{n,n'}\vfv_{m,k'}^{n'} d_{k'}^{a,n'}}_{\text{ICI (cross-user)}}
\notag\\
&\quad+
\underbrace{\sum_{n'=0}^{N-1}\sum_{m=1}^{M} \widetilde{\mathbf{G}}_{m,k}^{n,n'}\vfv_{m,k}^{n'} d_{k}^{a-1,n'}}_{\text{ISI (self)}}
+\underbrace{\sum_{n'=0}^{N-1}\sum_{k'\ne k}\sum_{m=1}^{M} \widetilde{\mathbf{G}}_{m,k}^{n,n'}\vfv_{m,k'}^{n'} d_{k'}^{a-1,n'}}_{\text{ISI (cross-user)}}
+\underbrace{\vfw_{k}^{a,n}}_{\text{noise}}.
\label{eq:final_rx_signal2}\tag{13}
\end{align}
\end{figure*}
Finally, by substituting \eqref{eq:tx_signal_time}, \eqref{eq:tx_signal_frequency}, and \eqref{eq:rx_signal_time} into \eqref{eq:rx_signal_frequency_mid}, the received signal can be expressed in the form of \eqref{eq:final_rx_signal2}.
For notational brevity, we further define the effective coefficients as
\begin{align}
\mathbf{G}_{m,k}^{n,n'} &\triangleq A_{nn'}(\delta_{m,k}^{s_k})\,\mH_{m,k}^n, \notag \\ 
\widetilde{\mathbf{G}}_{m,k}^{n,n'} &\triangleq B_{nn'}(\delta_{m,k}^{s_k})\,\mH_{m,k}^n. \notag
\end{align}
The received signal consists of four components: 
(i) the desired signal jointly transmitted from multiple satellites, 
(ii) MUI caused by signals intended for other UTs, 
(iii) ICI resulting from the loss of OFDM subcarrier orthogonality due to asynchronous reception within the same OFDM symbol, and 
(iv) ISI originating from previous OFDM symbols. 
It is noteworthy that the desired signal itself significantly contributes to interference, as asynchronous reception destroys OFDM symbol orthogonality and generates ISI.
Although ICI and ISI also arise from signals intended for other UTs, the interference caused by the desired signal is more dominant because it is beamformed toward the $k$-th UT.
It should be emphasized that the self-ICI and self-ISI terms pass through the same spatial channel $\mathbf{H}_{m,k}^{n}$ as the desired signal and differ from it only by the scalar leakage coefficients $A_{nn'}(\delta_{m,k}^{s_k})$ and $B_{nn'}(\delta_{m,k}^{s_k})$.
Hence, they cannot be suppressed by linear precoding, which operates on the spatial channel, and can only be avoided by keeping $\delta_{m,k}^{s_k}$ within the CP.
This observation motivates the proposed timing-aware association strategy.

\subsection{Achievable Downlink Throughput}
We assume that both the transmitter and the receiver have perfect CSI and that the Doppler shift is perfectly pre-compensated.
This allows us to isolate the effect of asynchronous reception without considering additional channel estimation errors.
The impact of imperfect CSI and residual Doppler errors is discussed in Section~\ref{sec_5}.D.
When the UT employs a linear combiner $\vfu \in \complex^{N_{\mathrm{rx}}\times1}$, the finally demodulated signal can be expressed as 
\setcounter{equation}{13}
\begin{equation}
    (\vfu^n_{k})^\mathrm{H}\vfy_k^{a,n}.
\end{equation}
The power of the desired signal, MUI, ICI, ISI, and noise is given by
\begin{gather}
P^{\mathrm{desired},n}_k
=
\Bigl|(\vfu_k^n)^\mathrm{H}
\sum_{m=1}^{M} \mathbf{G}_{m,k}^{n,n}\,\vfv_{m,k}^{n}\, d_{k}^{a,n}\Bigr|^2,
\notag\\[6pt]
P^{\mathrm{MUI},n}_k
=
\sum_{k' \ne k}
\Bigl|(\vfu_k^n)^\mathrm{H}
\sum_{m=1}^{M} \mathbf{G}_{m,k}^{n,n}\,\vfv_{m,k'}^{n}\, d_{k'}^{a,n}\Bigr|^2,
\notag\\[6pt]
P^{\mathrm{ICI},n}_k
=
\sum_{n' \ne n} \sum_{k'=1}^{K}
\Bigl|(\vfu_k^n)^\mathrm{H}
\sum_{m=1}^{M} \mathbf{G}_{m,k}^{n,n'}\,\vfv_{m,k'}^{n'}\, d_{k'}^{a,n'}\Bigr|^2,
\notag\\[6pt]
P^{\mathrm{ISI},n}_k
=
\sum_{n'=0}^{N-1} \sum_{k'=1}^{K}
\Bigl|(\vfu_k^n)^\mathrm{H}
\sum_{m=1}^{M} \widetilde{\mathbf{G}}_{m,k}^{n,n'}\,\vfv_{m,k'}^{n'}\, d_{k'}^{a-1,n'}\Bigr|^2,
\notag\\[6pt]
P^{\mathrm{noise},n}_k
=
\sigma_w^2 \bigl\Vert\vfu_k^n\bigr\Vert^2.
\label{eq:power_terms}
\end{gather}
Based on the power terms in \eqref{eq:power_terms}, the instantaneous signal-to-interference-plus-noise ratio (SINR) of the $k$-th UT on subcarrier $n$ is defined as
\begin{equation}
    \mathrm{SINR}_k^n
    =
    \frac{P^{\mathrm{desired},n}_k}{
    P^{\mathrm{MUI},n}_k
    + P^{\mathrm{ICI},n}_k
    + P^{\mathrm{ISI},n}_k
    + P^{\mathrm{noise},n}_k}.
    \label{eq:SINR_kn}
\end{equation}
Using this definition, the achievable downlink throughput of the $k$-th UT is given by
\begin{equation}
    R_k
    =
    \frac{N}{N+L_{\mathrm{CP}}}
    \sum_{n=0}^{N-1}
    B_{\mathrm{sc}}\log_2\bigl(1+\mathrm{SINR}_k^n\bigr),
    \label{eq:Rk_DL}
\end{equation}
where $B_{\mathrm{sc}}$ denotes the subcarrier bandwidth and the pre-log factor $\tfrac{N}{N+L_{\mathrm{CP}}}$ accounts for the loss of time-domain resources due to the CP overhead.

\begin{figure*}[!t]
\vspace{0mm}
    \centering
    \includegraphics[width=2\columnwidth]{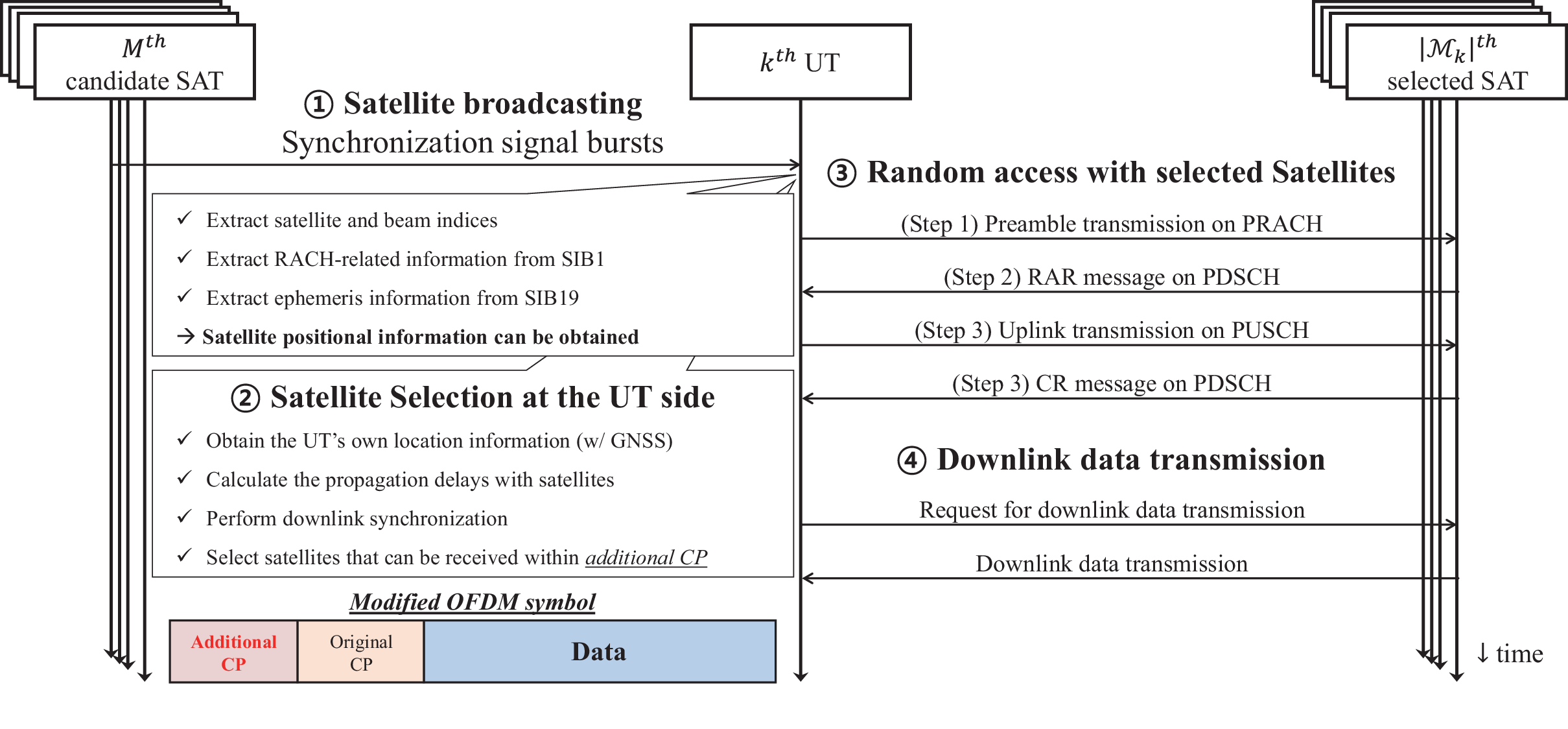}
    \vspace{-2mm}
    \caption{Overall framework of the proposed satellite association strategy.}
    \label{fig:overall_framework}
\end{figure*}

\section{Proposed Satellite Association Strategy}\label{sec_5}
\subsection{Overall Framework of the Proposed Strategy}
We design a strategy in which a UT associates only with satellites that induce acceptable levels of asynchronous reception. 
Leveraging the fact that timing offsets within the CP of an OFDM symbol preserve orthogonality and do not generate ISI, we allocate an additional CP $L_{\mathrm{CP}}^{\mathrm{add}}$ for multi-satellite connections. 
The UT then attempts to associate only with satellites whose timing offsets fall within this additional CP duration. 
The overall framework of the proposed strategy is illustrated in Fig.~\ref{fig:overall_framework}. The association strategy consists of the following steps.

\begin{figure}[!t]
    \centering
    \includegraphics[width=\columnwidth]{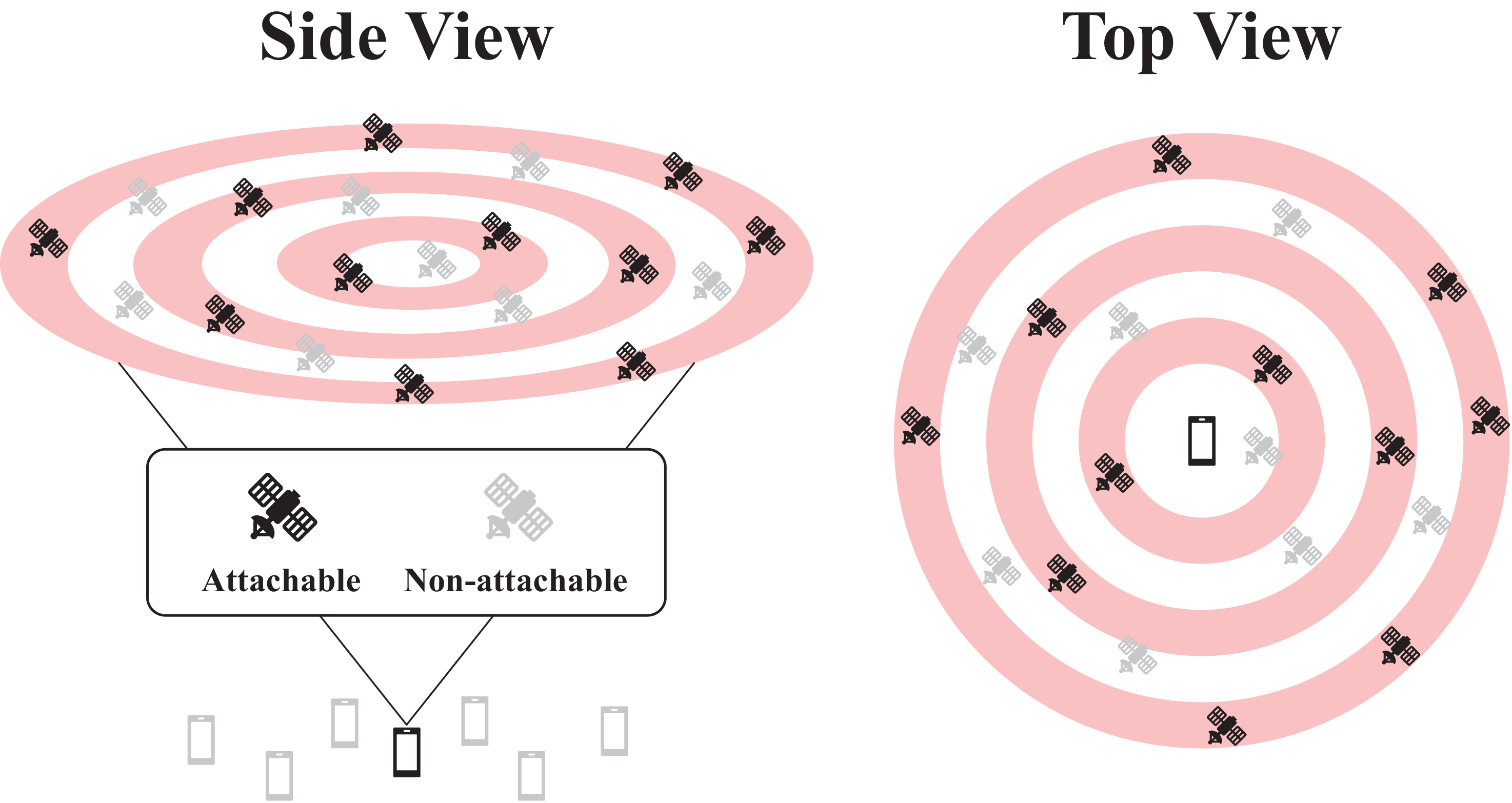}
    \caption{Illustration of the attachable region for a target UT.
    Satellites whose propagation delays, taken modulo the OFDM symbol length $N + L_{\mathrm{CP}} + L_{\mathrm{CP}}^{\mathrm{add}}$, fall within $L_{\mathrm{CP}}^{\mathrm{add}} + L_{\mathrm{margin}}$ are regarded as attachable, and the UT associates only with satellites in this region.}
    \label{fig:attachable_region}
\end{figure}

\begin{itemize}
    \item \textbf{Satellite broadcasting:} Satellites periodically broadcast synchronization signal bursts~\cite{38.821}. 
    From the physical broadcast channel (PBCH), the UT extracts the satellite index and beam index, which are then used to decode the system information blocks (SIBs) contained in the physical downlink shared channel (PDSCH).
    Among these, SIB1 provides the preamble configuration and random access channel (RACH) grant required for random access, while SIB19 contains the satellite ephemeris information. 
    Since ephemeris information includes the satellite's position and orbital parameters, the UT can obtain knowledge of both the current location and the future trajectory of the satellite.
    \item \textbf{Satellite selection at the UT side:} Since the UT is assumed to be equipped with GNSS functionality, it is aware of its own position. 
    Using this position information together with the satellite locations obtained from ephemeris data, the UT calculates the propagation delay between itself and each satellite. 
    The UT then performs downlink synchronization to align its timing with the satellites and to define the OFDM symbol boundary. 
    Based on the resulting symbol boundary, the UT defines the attachable region and considers only satellites located within this region, as illustrated in Fig.~\ref{fig:attachable_region}.
    Specifically, a satellite is selected as attachable if the timing offset $\delta_{m,k}^{s_k}$ in \eqref{eq:sample_offset}, with the symbol length extended to $N + L_{\mathrm{CP}} + L_{\mathrm{CP}}^{\mathrm{add}}$, satisfies $\delta_{m,k}^{s_k} \le L_{\mathrm{CP}}^{\mathrm{add}} + L_{\mathrm{margin}}$, i.e., if the remainder of the calculated propagation delay divided by the OFDM symbol length (including the CP) lies within $(L_{\mathrm{CP}}^{\mathrm{add}} + L_{\mathrm{margin}})$.
    Since symbol-level synchronization among satellites is maintained, the integer number of OFDM symbols contained in the propagation delay only shifts the symbol index at which the UT receives its data and is absorbed by adjusting the symbol index at the satellite, whereas the remaining sample-level offset $\delta_{m,k}^{s_k}$ cannot be compensated at the transmitter.
    Here, $L_{\mathrm{margin}}$ is introduced as a margin to account for satellite mobility and residual synchronization errors, and to ensure system stability.
    Because the propagation delay spans many OFDM symbols, the attachable region consists of multiple concentric rings, as illustrated in Fig.~\ref{fig:attachable_region}.
    \item \textbf{Random access with selected satellites:} Each UT attempts random access with all satellites selected in the previous step. 
    After completing the random access procedure, the UT is assigned IDs by all selected satellites, thereby establishing links with them.
    \item \textbf{Downlink data transmission:} When UTs have downlink data transmission demands, they send requests to the satellites whose signals can be received within the additional CP. 
    Upon receiving the requests, the satellites perform symbol synchronization and jointly transmit data to the UTs at the designated timing.
\end{itemize}
By performing satellite association through the above procedure, asynchronous interference among desired signals can be eliminated without requiring additional timing synchronization among satellites or extra beamforming. 
Once the UTs are attached to the satellites under the proposed strategy and begin receiving data services, the received signal in \eqref{eq:final_rx_signal2} can be rewritten as
\begin{align}
\vfy_{k}^{a,n}
&=
\underbrace{\sum_{m\in\mathcal{M}_k^{s_k}}
\mathbf{G}_{m,k}^{n,n}\,\vfv_{m,k}^{n}\, d_{k}^{a,n}}_{\scriptstyle\text{desired signal}} \notag\\
&\quad+
\underbrace{\sum_{k' \ne k}\sum_{m\in\mathcal{M}_{k'}^{s_{k'}}}
\mathbf{G}_{m,k}^{n,n}\,\vfv_{m,k'}^{n}\, d_{k'}^{a,n}}_{\scriptstyle\text{MUI}} \notag\\
&\quad+
\underbrace{\sum_{n' \ne n}\sum_{k' \ne k}\sum_{m\in\mathcal{M}_{k'}^{s_{k'}}}
\mathbf{G}_{m,k}^{n,n'}\,\vfv_{m,k'}^{n'}\, d_{k'}^{a,n'}}_{\scriptstyle\text{ICI (cross-user)}} \notag\\
&\quad+
\underbrace{\sum_{n'=0}^{N-1}\sum_{k' \ne k}\sum_{m\in\mathcal{M}_{k'}^{s_{k'}}}
\widetilde{\mathbf{G}}_{m,k}^{n,n'}\,\vfv_{m,k'}^{n'}\, d_{k'}^{a-1,n'}}_{\scriptstyle\text{ISI (cross-user)}}
+\underbrace{\vfw_{k}^{a,n}}_{\scriptstyle\text{noise}}.
\label{eq:final_rx_signal_prop}
\end{align}
\noindent where $\mathcal{M}_k^{s_k}$ denotes the set of satellites associated with the $k$-th UT.
The power terms in \eqref{eq:power_terms} are modified as follows (the noise term is unchanged and thus omitted):
{\interdisplaylinepenalty=10000
\begin{gather}
P^{\mathrm{desired},n}_{k,\mathrm{mod}}
=
\Bigl|(\vfu_k^n)^\mathrm{H}
\sum_{m\in{\mathcal{M}_k^{s_k}}}
\mathbf{G}_{m,k}^{n,n}\,\vfv_{m,k}^{n}\, d_{k}^{a,n}\Bigr|^2
\notag\\[6pt]
P^{\mathrm{MUI},n}_{k,\mathrm{mod}}
=
\sum_{k' \ne k}
\Bigl|(\vfu_k^n)^\mathrm{H}
\sum_{m\in\mathcal{M}_{k'}^{s_{k'}}}
\mathbf{G}_{m,k}^{n,n}\,\vfv_{m,k'}^{n}\, d_{k'}^{a,n}\Bigr|^2
\notag\\[6pt]
P^{\mathrm{ICI},n}_{k,\mathrm{mod}}
=
\sum_{n' \ne n}\sum_{k'\ne k}
\Bigl|(\vfu_k^n)^\mathrm{H}
\sum_{m\in\mathcal{M}_{k'}^{s_{k'}}}
\mathbf{G}_{m,k}^{n,n'}\,\vfv_{m,k'}^{n'}\, d_{k'}^{a,n'}\Bigr|^2
\notag\\[6pt]
P^{\mathrm{ISI},n}_{k,\mathrm{mod}}
=
\sum_{n'=0}^{N-1}\sum_{k'\ne k}
\Bigl|(\vfu_k^n)^\mathrm{H}
\sum_{m\in\mathcal{M}_{k'}^{s_{k'}}}
\widetilde{\mathbf{G}}_{m,k}^{n,n'}\,\vfv_{m,k'}^{n'}\, d_{k'}^{a-1,n'}\Bigr|^2
\label{eq:modified_power_terms}
\end{gather}}
Then the total throughput is given by
\begin{align}
R_k
&=
\frac{N}{N+L_{\mathrm{CP}}+L_{\mathrm{CP}}^{\mathrm{add}}}
\sum_{n=0}^{N-1}
B_{\mathrm{sc}}\log_2\!\Biggl(1+\frac{P^{\mathrm{desired},n}_{k,\mathrm{mod}}}{\Psi_{k,n}}\Biggr),
\label{eq:modified_total_spectral_efficiency}\\
\Psi_{k,n}
&\triangleq
P^{\mathrm{MUI},n}_{k,\mathrm{mod}}
+P^{\mathrm{ICI},n}_{k,\mathrm{mod}}
+P^{\mathrm{ISI},n}_{k,\mathrm{mod}}
+P^{\mathrm{noise},n}_{k}.
\nonumber
\end{align}
An important distinction from \eqref{eq:Rk_DL} is that no asynchronous reception occurs among the desired signals; hence, the significant ICI and ISI contributions from the desired signal are eliminated.
While increasing the additional CP length for multi-LEO satellite association allows a UT to connect to a larger number of satellites, it also reduces time resource efficiency and can thus impact the total spectral efficiency. 
Therefore, the trade-off between these two factors must be carefully considered in system parameter design.

\subsection{Spectral Efficiency Analysis with Respect to CP Length}
As discussed earlier, the CP length plays an important role because it directly affects both the power gain and the time-resource efficiency in the proposed satellite association strategy. 
We analyze how the CP length influences the spectral efficiency achieved by the proposed association strategy.

In this subsection, we focus only on the additional CP required for satellite association and omit the conventional CP for simplicity, so that the OFDM symbol length is $L \triangleq N + L_{\mathrm{CP}}^{\mathrm{add}}$, and set $L_{\mathrm{margin}} = 0$.
Furthermore, downlink synchronization is performed randomly, i.e., independently of the satellite positions, and the set of satellites selected for the $k$-th UT is denoted by $\mathcal{M}_k$.
To simplify the SINR expression, we define the aggregate interference-plus-noise power on subcarrier $n$ for the $k$-th UT as
\begin{align}
\Gamma_{k,n}
&\triangleq
P^{\mathrm{MUI},n}_{k,\mathrm{mod}}
+P^{\mathrm{ICI},n}_{k,\mathrm{mod}}
+P^{\mathrm{ISI},n}_{k,\mathrm{mod}}
+P^{\mathrm{noise},n}_{k}.
\label{eq:Gamma_def}
\end{align}
Then, the ergodic spectral efficiency corresponding to \eqref{eq:modified_total_spectral_efficiency} can be expressed as
\begin{align}
\mathbb{E}[R_k]
&=\mathbb{E}\!\Bigg[
\frac{NB_{\mathrm{sc}}}{N+L_{\mathrm{CP}}^{\mathrm{add}}}
\sum_{n=0}^{N-1}
\log_2\!\left(
1+\frac{P^{\mathrm{desired},n}_{k,\mathrm{mod}}}{\Gamma_{k,n}}
\right)
\Bigg] \notag\\[4pt]
&\overset{\text{(a)}}{\approx}
\frac{N^2B_{\mathrm{sc}}}{N+L_{\mathrm{CP}}^{\mathrm{add}}}\,
\mathbb{E}\!\left[
\log_2\!\left(
1+\frac{P^{\mathrm{desired},n}_{k,\mathrm{mod}}}{\Gamma_{k,n}}
\right)
\right] \notag\\[4pt]
&\overset{\text{(b)}}{\approx}
\frac{N^2B_{\mathrm{sc}}}{N+L_{\mathrm{CP}}^{\mathrm{add}}}\,
\log_2\!\left(
1+\frac{\mathbb{E}\!\left[P^{\mathrm{desired},n}_{k,\mathrm{mod}}\right]}
{\mathbb{E}\!\left[\Gamma_{k,n}\right]}
\right).
\label{eq:ergodic_rate_final}
\end{align}
The approximation in (a) follows from the fact that the SINR term is identically distributed across subcarriers, so the summation over $N$ subcarriers can be approximated by $N$ times the per-subcarrier expectation for sufficiently large $N$.
The approximation in (b) is justified because both the numerator and denominator of the SINR term are sums of nonnegative random variables, which allows us to apply Lemma~1 in \cite{JSTSP14_TP} to obtain
$\mathbb{E}\!\left[\log_2\!\left(1+\frac{X}{Y}\right)\right]\approx \log_2\!\left(1+\frac{\mathbb{E}[X]}{\mathbb{E}[Y]}\right)$.
Consequently, the spectral efficiency can be characterized using only the expectations of the desired-signal power and the interference-plus-noise power components.

For analytical simplicity, we assume that all UTs are equipped with a single antenna, and we adopt the free-space path loss model~\cite{TVT25_FSPL}. 
Under this assumption, the channel can be expressed as
\begin{equation}
    \mH_{m,k}^{n} = \frac{c}{4\pi r_{m,k}f_{\mathrm{c}}} \,\cdot \alpha_{m,k}^n\,\left\{\va^{\mathrm{SAT},n}\left(\theta^{(\mathrm{SAT}_m)}_{m,k}, \phi^{(\mathrm{SAT}_m)}_{m,k}\right)\right\}^\mathrm{H},
    \label{eq:simplified_channel}
\end{equation}
where $r_{m,k}$ denotes the distance between the $m$-th satellite and the $k$-th UT, and $f_{\mathrm{c}}$ is the center frequency.
For precoding, we employ a maximum ratio transmission (MRT) precoder which incorporates a phase compensation term induced by the propagation delay.
The precoder is given by
\begin{equation}
\vfv_{m,k}^{n}
= e^{j\frac{2\pi n \delta_{m,k}}{N}}
\sqrt{P_{m,k}}
\left(
    \frac{\mH_{m,k}^{n}}
    {\left\|\mH_{m,k}^{n}\right\|}
\right)^\mathrm{H}. \label{eq:analysis_precoder}
\end{equation}
The transmit power $P_{m,k}$ is assumed to be maximized under the power constraint defined by the received power flux density (PFD) regulations of the International Telecommunication Union (ITU)~\cite{ITU}. 
Accordingly, $P_{m,k}$ can be expressed as follows:
\begin{equation}
P_{m,k}
= 4\pi r_{m,k}^{2}
\cdot
\frac{\mathrm{PFD}}{\lvert\mathcal{M}_k\rvert}. \label{eq:normalized_power}
\end{equation}
Based on these conditions, we derive the expectation of each power term.

First, the expectation of the desired power is given by
\begin{align}\nonumber
\mathbb{E}\!\left[P_{k,\mathrm{mod}}^{\mathrm{desired},n}\right]
&= \mathbb{E}\!\left[
\left|
\sum_{m\in\mathcal{M}_k}
\mathbf{G}_{m,k}^{n,n}\,\vfv_{m,k}^n d_{k}^{a,n}
\right|^{2}
\right] \\\nonumber
&= \mathbb{E}\!\left[
\left|
\sum_{m\in\mathcal{M}_k}
\sqrt{\frac{\mathrm{PFD}_0}{|\mathcal{M}_k|}}\,
\lvert \alpha_{m,k} \rvert
\right|^{2}
\right] \\
&= \mathbb{E}\!\left[
\frac{\mathrm{PFD}_0}{|\mathcal{M}_k|}
\left(
\sum_{m\in\mathcal{M}_k} \lvert \alpha_{m,k} \rvert
\right)^{2}
\right], \label{eq:desired_power_expect}
\end{align}
where $\mathrm{PFD}_0 = \mathrm{PFD} \cdot c^{2}/(4\pi f_{\mathrm{c}}^{2})$. 
Given $|\mathcal{M}_k|$ and defining $\mu_{1} \triangleq \mathbb{E}[|\alpha|]$ and $\mu_{2} \triangleq \mathbb{E}[|\alpha|^{2}]$, the expectation of the desired power becomes
\begin{equation}
\mathbb{E}\!\bigl[P_{k,\mathrm{mod}}^{\mathrm{desired},n}\bigm|\lvert\mathcal{M}_k\rvert\bigr]
= \mathrm{PFD}_0\bigl(\mu_2 + (\lvert\mathcal{M}_k\rvert - 1)\mu_1^{\,2}\bigr). \label{eq:desired_power_cond}
\end{equation}
Since the satellite positions are i.i.d. uniformly distributed over the coverage region \cite{okati2020downlink,park2023tractable} and the synchronization point is independent of them, the indicators $\ind{\delta_{m,k}^{s_k} \le L_{\mathrm{CP}}^{\mathrm{add}}}$, $m = 1, \ldots, M$, are i.i.d., and hence $|\mathcal{M}_k|$ follows a binomial distribution as $|\mathcal{M}_k| \sim \mathrm{Binomial}(M, p_{\mathrm{CP}})$, where $p_{\mathrm{CP}} = \Pr\{\delta_{m,k}^{s_k} \le L_{\mathrm{CP}}^{\mathrm{add}}\}$.
As the propagation delay spans many OFDM symbols, $\delta_{m,k}^{s_k}$ is approximately uniformly distributed over $\{0, \ldots, L-1\}$, which yields $p_{\mathrm{CP}} = L_{\mathrm{CP}}^{\mathrm{add}} / (N + L_{\mathrm{CP}}^{\mathrm{add}})$, i.e., the ratio of the attachable window to the symbol length.
Consequently, the desired power can be expressed as
\begin{align}
\mathbb{E}\!\left[P_{k,\mathrm{mod}}^{\mathrm{desired},n}\right]
&= \mathbb{E}_{|\mathcal{M}_k|}\!\left[
\mathrm{PFD}_0\bigl(\mu_2 + (|\mathcal{M}_k|-1)\mu_1^2\bigr)
\right] \notag \\
&= \mathrm{PFD}_0\bigl(\mu_2 + (\mathbb{E}[|\mathcal{M}_k|]-1)\mu_1^2\bigr) \notag \\
&= \mathrm{PFD}_0\bigl(\mu_2 + (M\cdot p_{\mathrm{CP}}-1)\mu_1^2\bigr).  \label{eq:desired_power_binomial}
\end{align}
Since the small-scale fading is normalized and the channel is LoS-dominant, we approximate $\mu_{1}=\mu_{2}=1$ \cite{Goldsmith2005Wireless,Molisch2011Wireless}.
With this approximation, the closed-form expression of the desired power is given by
\begin{equation}
   \mathbb{E}\!\left[P_{k,\mathrm{mod}}^{\mathrm{desired},n}\right] = \mathrm{PFD}_0\cdot M\cdot \frac{L_{\mathrm{CP}}^{\mathrm{add}}}{N+L_{\mathrm{CP}}^{\mathrm{add}}}. \label{eq:desired_power_closed}
\end{equation}

Next, we examine the interference power. The expectation of the MUI power is given by
\begin{align}
\mathbb{E}\!\left[P_{k,\mathrm{mod}}^{\mathrm{MUI},n}\right]
&= \mathbb{E}\!\Biggl[
\sum_{k'\neq k}\Bigl|\sum_{m\in\mathcal{M}_{k'}}
\mathbf{G}_{m,k}^{n,n}\,\vfv_{m,k'}^n d_{k'}^{a,n}
\Bigr|^{2}
\Biggr] \notag\\
&= \mathbb{E}\!\Biggl[
\sum_{k'\neq k}\Bigl|\sum_{m\in\mathcal{M}_{k'}}
\frac{c}{4\pi f_{\mathrm{c}} r_{m,k}}\sqrt{P_{m,k'}}
\notag\\
&\quad\qquad\qquad\qquad\cdot
|\alpha_{m,k}|\,|\rho_m(k,k')|\,
e^{j\phi_{m,k,k'}}
\Bigr|^{2}
\Biggr] \notag\\
&\approx \sum_{k'\neq k}\sum_{m\in\mathcal{M}_{k'}}
\mathbb{E}\!\Biggl[
\mathrm{PFD}_0\,
\frac{r_{m,k'}^{2}}{r_{m,k}^{2}}\,
\frac{|\alpha_{m,k}|^{2}}{|\mathcal{M}_{k'}|}
\notag\\
&\quad\qquad\qquad\qquad\cdot
|\rho_m(k,k')|^{2}\,
\Bigl(\frac{N-\Delta_{m,k}}{N}\Bigr)^{2}
\Biggr],
\label{eq:Emui_mod}
\end{align}
\noindent where $\phi_{m,k,k'}\triangleq -\tfrac{2\pi n}{N}\delta_{m,k}+\arg(\rho_m(k,k'))$.
We denote $\rho_{m}(k,k')$ as the correlation between the array responses of the $k$-th and $k'$-th UTs with respect to the $m$-th satellite.
The approximation in the above expression is valid because both the sample offset $\delta$ and the phase of the array-response correlation are uniformly distributed and independent across satellites, given that the satellites and UTs are uniformly deployed.
This independence concerns the remainders of the propagation delays rather than the delays themselves. Since the symbol length corresponds to about 17~km of range while the ranges to the visible satellites span more than a thousand kilometers, the remainders are scrambled across satellites even when the satellite positions are correlated, as in structured constellations.
Similar to the desired power, the expectation of the MUI power can be obtained by conditioning on $|\mathcal{M}_{k'}|$, which yields
\begin{align}\nonumber
\mathbb{E}\!\left[P^{\mathrm{MUI},n}_{k,\mathrm{mod}}\right]
&= (K-1)\cdot
\mathbb{E}\!\Biggl[
\mathrm{PFD}_0 \cdot \frac{r_{m,k'}^{2}}{r_{m,k}^{2}} \cdot
|\alpha_{m,k}|^{2}
\nonumber\\
&\quad\qquad\qquad\qquad\cdot
|\rho_m(k,k')|^{2} \cdot
\Bigl(\frac{N-\Delta_{m,k}}{N}\Bigr)^{2}
\Biggr].
\end{align}
For analytical tractability, we adopt standard large-scale/small-scale channel modeling and assume that the path loss, small-scale fading, array response, and residual delay are mutually independent~\cite{Goldsmith2005Wireless,Molisch2011Wireless}. We focus on LoS-dominant satellite--UT links, where most of the received power comes from a deterministic LoS component and the remaining diffuse multipath component is modeled as wide-sense stationary and uncorrelated scattering (WSSUS).
Under these assumptions, MUI can be expressed as 
\begin{align}
\mathbb{E}\!\left[P^{\mathrm{MUI},n}_{k,\mathrm{mod}}\right]
&= (K-1)\cdot \mathrm{PFD}_0 \cdot
\mathbb{E}\!\left[\frac{r_{m,k'}^{2}}{r_{m,k}^{2}}\right]
\cdot \mathbb{E}\!\left[|\alpha_{m,k}|^{2}\right]
\notag\\
&\quad\cdot \mathbb{E}\!\left[|\rho_m(k,k')|^{2}\right]
\cdot \mathbb{E}\!\left[\left(\frac{N-\Delta_{m,k}}{N}\right)^{2}\right]
\notag\\
&\approx (K-1)\cdot \mathrm{PFD}_0 \cdot
\mathbb{E}\!\left[|\rho_m(k,k')|^{2}\right]
\!\cdot\! \frac{N/3 + L_{\mathrm{CP}}^{\mathrm{add}}}{N + L_{\mathrm{CP}}^{\mathrm{add}}}
\notag\\
&\ge (K-1)\cdot \mathrm{PFD}_0 \cdot
\frac{1}{N_{\mathrm{tx}}}
\cdot \frac{N/3 + L_{\mathrm{CP}}^{\mathrm{add}}}{N + L_{\mathrm{CP}}^{\mathrm{add}}}.
\label{eq:Emui_bound}
\end{align}
The lower bound corresponds to the case where the angles of departure (AoD) and arrival (AoA) of each UT are isotropically and uniformly distributed, resulting in an array-response correlation of $1/N_{\mathrm{tx}}$.

The ICI and ISI terms can be derived in a similar manner to the MUI term, leading to
{\interdisplaylinepenalty=10000
\begin{align}
\mathbb{E}\!\left[P_{k,\mathrm{mod}}^{\mathrm{ICI},n}\right]
&= \mathbb{E}\!\Biggl[
\sum_{n' \neq n} \sum_{k' \neq k} \Bigl|\sum_{m \in \mathcal{M}_{k'}}
\mathbf{G}_{m,k}^{n,n'}\,\vfv_{m,k'}^{n'} d_{k'}^{a,n'}
\Bigr|^{2}
\Biggr] \notag \\
&\approx (K-1)\cdot \mathrm{PFD}_0 \cdot
\mathbb{E}\!\left[\frac{r_{m,k'}^{2}}{r_{m,k}^{2}}\right]
\cdot \mathbb{E}\!\left[|\alpha_{m,k}|^{2}\right]
\notag\\
&\quad\cdot \mathbb{E}\!\left[|\rho_{m}(k,k')|^{2}\right]
\cdot \mathbb{E}\!\Bigl[\sum_{n' \neq n} \bigl|A_{nn'}(\delta_{m,k})\bigr|^{2}\Bigr]
\notag\\
&\ge (K-1)\cdot \mathrm{PFD}_0 \cdot
\frac{1}{N_{\mathrm{tx}}}\cdot \frac{1}{6}\cdot
\frac{N}{N + L_{\mathrm{CP}}^{\mathrm{add}}},
\label{eq:Eici_bound}
\end{align}}
\begin{align}
\mathbb{E}\!\left[P_{k}^{\mathrm{ISI},n}\right]
&= \mathbb{E}\!\Biggl[
\sum_{n'=0}^{N-1} \sum_{k' \neq k} \Bigl| \sum_{m \in \mathcal{M}_{k'}}
\widetilde{\mathbf{G}}_{m,k}^{n,n'}\,\vfv_{m,k'}^{n'} d_{k'}^{a-1,n'}
\Bigr|^{2}
\Biggr] \notag \\
&\approx (K-1)\cdot \mathrm{PFD}_0 \cdot
\mathbb{E}\!\left[\frac{r_{m,k'}^{2}}{r_{m,k}^{2}}\right]
\cdot \mathbb{E}\!\left[|\alpha_{m,k}|^{2}\right]
\notag\\
&\quad\cdot \mathbb{E}\!\left[|\rho_{m}(k,k')|^{2}\right]
\cdot \mathbb{E}\!\left[\sum_{n'=0}^{N-1} \bigl|B_{nn'}(\delta_{m,k})\bigr|^{2}\right]
\notag \\
&\ge (K-1)\cdot \mathrm{PFD}_0 \cdot
\frac{1}{N_{\mathrm{tx}}}\cdot \frac{1}{2}\cdot
\frac{N}{N + L_{\mathrm{CP}}^{\mathrm{add}}}.
\label{eq:Eisi_bound}
\end{align}

Finally, the average throughput can be expressed as
\begin{align}
\mathbb{E}[R_k]
&= \frac{N^{2}B_{\mathrm{sc}}}{N + L_{\mathrm{CP}}^{\mathrm{add}}}
\log_{2}\!\Biggl(
1 +
\frac{\substack{\mathrm{PFD}_0 \, M \,
\frac{L_{\mathrm{CP}}^{\mathrm{add}}}{N + L_{\mathrm{CP}}^{\mathrm{add}}}}}
{\substack{
(K-1)\,\mathrm{PFD}_0 \,\mathbb{E}\left[|\rho_m(k,k')|^{2}\right] + \sigma_w^2
}}
\Biggr)
\notag\\
&\le
\frac{N^{2}B_{\mathrm{sc}}}{N + L_{\mathrm{CP}}^{\mathrm{add}}}
\log_{2}\!\Biggl(
1 +
\frac{\substack{\mathrm{PFD}_0 \, M \,
\frac{L_{\mathrm{CP}}^{\mathrm{add}}}{N + L_{\mathrm{CP}}^{\mathrm{add}}}}}
{\substack{
(K-1)\,\mathrm{PFD}_0 \,\frac{1}{N_{\mathrm{tx}}} + \sigma_w^2
}}
\Biggr).
\label{eq:spectral_efficiency_analysis}
\end{align}
As the CP length increases, the number of satellites that can be associated with each UT grows, resulting in a proportional power gain. 
Since the MUI, ICI, and ISI components are combined in a non-coherent manner, their magnitudes do not scale with the number of satellites. 
The MUI increases with the CP length because it is generated by signals arriving within the CP duration, whereas the ICI and ISI decrease as the CP length increases because they originate from signals arriving outside the CP duration. 
These opposing effects cancel each other, leading to interference terms that are effectively independent of the CP length.
In contrast, the pre-log term decreases with the CP length due to the increased redundancy, which reduces the time-resource efficiency. 
Consequently, and consistent with intuition, increasing the CP length introduces a fundamental trade-off between power gain and time-resource efficiency.
Although the CP length that maximizes \eqref{eq:spectral_efficiency_analysis} is not available in closed form, the optimum can be characterized by a single parameter as follows.

Let $\eta \triangleq L_{\mathrm{CP}}^{\mathrm{add}}/(N + L_{\mathrm{CP}}^{\mathrm{add}})$ denote the fraction of the additional CP in the modified OFDM symbol, and let $\gamma_M \triangleq \mathrm{PFD}_0 M/\big((K-1)\mathrm{PFD}_0/N_{\mathrm{tx}} + \sigma_w^2\big)$ denote the SINR obtained when all $M$ satellites are coherently combined.
Then \eqref{eq:spectral_efficiency_analysis} is written as $N B_{\mathrm{sc}}(1-\eta)\log_2(1+\gamma_M \eta)$, and the optimal fraction $\eta^\ast$ is the unique solution of
\begin{equation}
\frac{(1-\eta^\ast)\,\gamma_M}{1+\gamma_M \eta^\ast} = \ln\big(1+\gamma_M \eta^\ast\big),
\label{eq:optimal_CP}
\end{equation}
which is a scalar equation solved by a one-dimensional search, with the optimal additional CP length $L_{\mathrm{CP}}^{\mathrm{add}\ast} = N \eta^\ast/(1-\eta^\ast)$.
The deployment parameters affect $L_{\mathrm{CP}}^{\mathrm{add}\ast}$ only through $\gamma_M$, and the FFT size scales $L_{\mathrm{CP}}^{\mathrm{add}\ast}$ linearly.
Accordingly, a longer CP is preferred in sparse constellations or at low signal-to-noise ratio (SNR), where $\eta^\ast$ approaches $1/2$, while a shorter CP is preferred in dense constellations or at high SNR; for example, $\eta^\ast$ is 0.34 and 0.23 for $\gamma_M$ of 10 and 20~dB.

\subsection{Downlink Synchronization}\label{sec3-2-2}
In the proposed satellite association strategy, downlink synchronization plays a critical role because it determines the number of satellites that can be associated with each UT.
If downlink synchronization is performed arbitrarily, a UT can on average associate with $\Bigl(M \times \frac{L_{\mathrm{CP}}^{\mathrm{add}}}{N+L_{\mathrm{CP}}+L_{\mathrm{CP}}^{\mathrm{add}}}\Bigr)$ satellites, corresponding to the ratio of the additional CP length to the total symbol length. 
However, the instantaneous number of associated satellites may be either smaller or larger than this average. 
When the number exceeds the average, additional power gain can be achieved, which is advantageous. 
Conversely, if fewer satellites are associated, the desired signal power becomes insufficient while the interference from other UTs remains unchanged, leading to reduced spectral efficiency. 
Therefore, each UT should perform downlink synchronization in a way that maximizes the number of associated satellites, which can be formulated as the following optimization problem.
\begin{equation}
\begin{aligned}
s_k^\star
&= \arg\max_{s_k \in \{0,1,\dots,N-1\}} \bigl|\mathcal{M}_k^{s_k}\bigr|
\\[6pt]
\text{s.t.}\quad
&\delta_{m,k}^{s_k}
= \Bigl(\Bigl\lfloor\tfrac{\tau_{m,k}}{T_{\mathrm{s}}}\Bigr\rfloor - s_k\Bigr)
\bmod \bigl(N+L_{\mathrm{CP}}+L_{\mathrm{CP}}^{\mathrm{add}}\bigr), \\
&\mathcal{M}_k^{s_k}
= \bigl\{\, m \in \mathcal{M} : \delta_{m,k}^{s_k} \le L_{\mathrm{CP}}^{\mathrm{add}} \,\bigr\},
\end{aligned}
\label{eq:downlink_synchronization}
\end{equation}
where $\mathcal{M} \triangleq \{1, \ldots, M\}$.
Problem \eqref{eq:downlink_synchronization} can be solved exactly with low complexity as follows.
Let $L_{\mathrm{sym}} \triangleq N + L_{\mathrm{CP}} + L_{\mathrm{CP}}^{\mathrm{add}}$ and $d_{m,k} \triangleq \lfloor \tau_{m,k}/T_{\mathrm{s}} \rfloor \bmod L_{\mathrm{sym}}$.
The $m$-th satellite is attachable if and only if $s_k \in \{d_{m,k} - L_{\mathrm{CP}}^{\mathrm{add}}, \ldots, d_{m,k}\}$ modulo $L_{\mathrm{sym}}$, i.e., each satellite corresponds to an interval of length $L_{\mathrm{CP}}^{\mathrm{add}} + 1$ on the circular time axis of length $L_{\mathrm{sym}}$.
Hence, $|\mathcal{M}_k^{s_k}|$ is the number of intervals containing $s_k$, and \eqref{eq:downlink_synchronization} is to find the point in $\{0, \ldots, N-1\}$ covered by the largest number of intervals.
Since $|\mathcal{M}_k^{s_k}|$ increases only at the starting points $a_{m,k} \triangleq (d_{m,k} - L_{\mathrm{CP}}^{\mathrm{add}}) \bmod L_{\mathrm{sym}}$ of the intervals, the optimum is attained either at $s_k = 0$ or at one of the starting points within the search range.
Algorithm~\ref{alg:downlink_sync} sorts the $2M$ boundary points and sweeps over them once while counting the covering intervals, which yields the global optimum of \eqref{eq:downlink_synchronization} with complexity $\mathcal{O}(M \log M)$.
The algorithm is executed once per UT whenever the satellite positions are updated from the ephemeris information.

\begin{algorithm}[!t]
\caption{Downlink synchronization for the $k$-th UT}
\begin{algorithmic}[1]
\Require $\{\tau_{m,k}\}_{m \in \mathcal{M}}$, $T_{\mathrm{s}}$, $N$, $L_{\mathrm{sym}}$, $L_{\mathrm{CP}}^{\mathrm{add}}$
\Ensure $s_k^\star$
\State $\mathcal{E} \gets \emptyset$, $c \gets 0$
\For{$m \in \mathcal{M}$}
    \State $d_{m,k} \gets \lfloor \tau_{m,k}/T_{\mathrm{s}} \rfloor \bmod L_{\mathrm{sym}}$
    \State $a_{m,k} \gets (d_{m,k} - L_{\mathrm{CP}}^{\mathrm{add}}) \bmod L_{\mathrm{sym}}$
    \State $\mathcal{E} \gets \mathcal{E} \cup \{(a_{m,k}, +1),\, (d_{m,k} + 1, -1)\}$
    \If{$a_{m,k} > d_{m,k}$} \State $c \gets c + 1$ \Comment{interval wraps around and covers $0$} \EndIf
\EndFor
\State Sort $\mathcal{E}$ in ascending order of position
\State $c^\ast \gets c$, $s_k^\star \gets 0$
\For{each distinct position $p \le N - 1$ in $\mathcal{E}$}
    \State $c \gets c + \sum_{(p, \Delta) \in \mathcal{E}} \Delta$
    \If{$c > c^\ast$} \State $c^\ast \gets c$, $s_k^\star \gets p$ \EndIf
\EndFor
\State \Return $s_k^\star$
\end{algorithmic} \label{alg:downlink_sync}
\end{algorithm}

\begin{figure}[t]
    \centering
    \includegraphics[width=1\columnwidth]{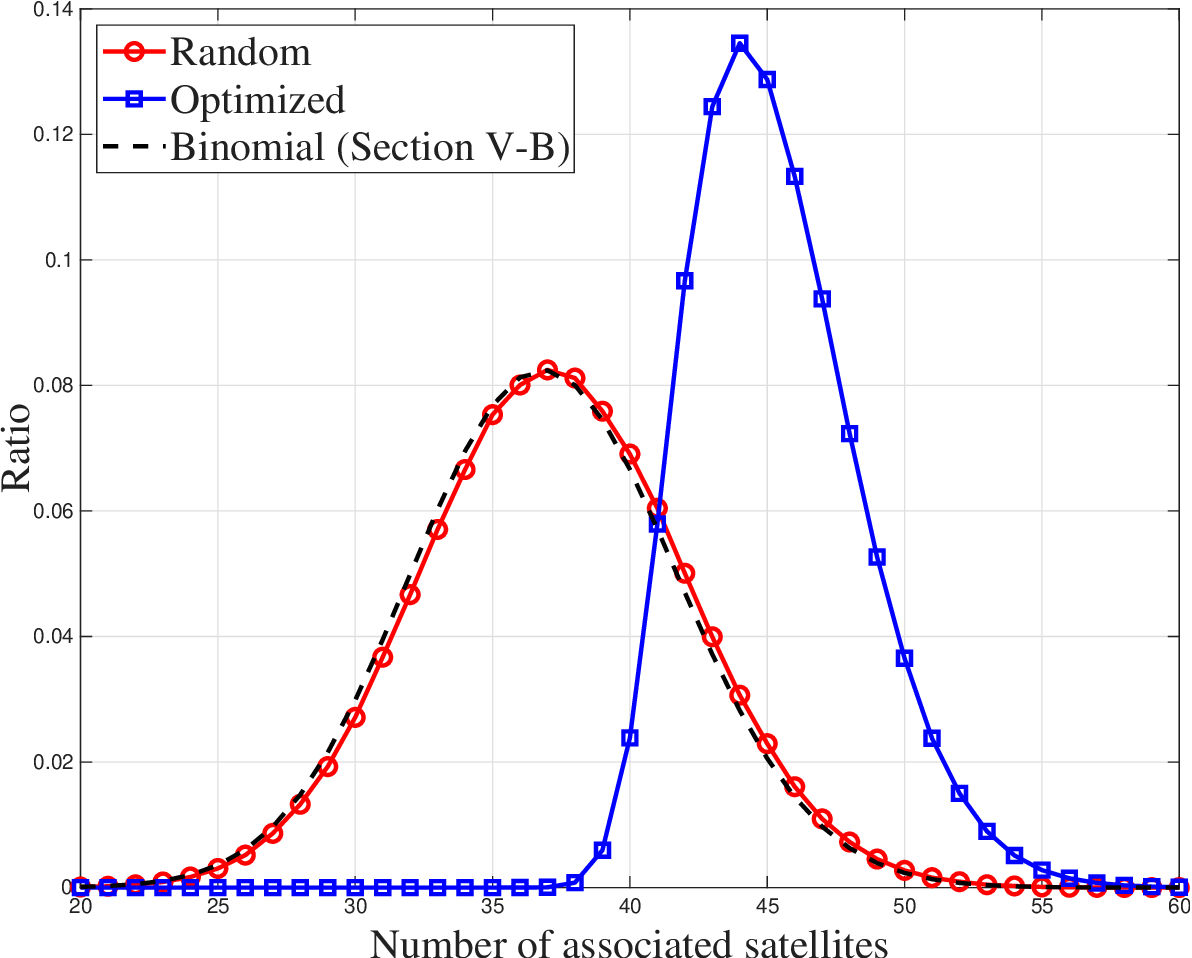}
    \caption{Normalized histogram of the number of satellites associated with a UT, represented as line plots for clarity. The dashed line denotes the binomial distribution in Section~\ref{sec_5}.B.}
    \label{fig:attach_example}
\end{figure}

By performing downlink synchronization through the proposed optimization, a UT can maximize the number of satellites associated in each instantaneous scenario. 
For example, consider a system with a minimum elevation angle of $10^\circ$, where 100 satellites at an altitude of 600\,km are uniformly distributed within the coverage region of a UT and the additional CP length is set to 600. 
The normalized histogram of the number of associated satellites is presented in Fig.~\ref{fig:attach_example}, with line plots used to improve visual clarity.
When downlink synchronization is performed randomly, the number of associated satellites ranges from about 20 to 55, and its distribution closely follows the binomial distribution in Section~\ref{sec_5}.B, which validates the independence of the sample offsets across satellites.
In contrast, when synchronization is performed according to \eqref{eq:downlink_synchronization}, the number of associated satellites ranges from about 38 to 58, enabling a UT to stably associate with a much larger set of satellites.
Consequently, the proposed downlink synchronization not only maximizes the number of associated satellites but also reduces the variation across UTs, thereby enhancing the overall spectral efficiency.

\subsection{Impact of Imperfect CSI and Residual Doppler}
The above analysis assumes perfect CSI and perfect Doppler compensation.
Under the free-space path loss model in \eqref{eq:simplified_channel}, the array response and the path loss are determined by the geometry, and the CSI to be estimated reduces to the carrier phase of each satellite.
We therefore model imperfect CSI as a residual phase error $\zeta_m$ at the $m$-th satellite, assumed i.i.d. with zero mean.
With this error, the desired signal in \eqref{eq:desired_power_expect} becomes $\sum_{m\in\mathcal{M}_k}|\alpha_{m,k}|e^{j\zeta_m}$, and following \eqref{eq:desired_power_cond}--\eqref{eq:desired_power_closed}, the desired power is
\begin{equation}
\mathbb{E}\big[ P_{k,\mathrm{mod}}^{\mathrm{desired},n} \big] = \mathrm{PFD}_0 \big( 1 + (M p_{\mathrm{CP}} - 1)\kappa \big),
\label{eq:desired_power_error}
\end{equation}
where $\kappa \triangleq |\mathbb{E}[e^{j\zeta}]|^2$, which equals $e^{-\sigma_\zeta^2}$ for Gaussian phase errors with standard deviation $\sigma_\zeta$.
The interference powers in \eqref{eq:Emui_mod}--\eqref{eq:Eisi_bound} are unchanged, since they are already averaged over independent phases, and the association in \eqref{eq:downlink_synchronization} does not involve CSI.
The phase error thus scales the coherent power gain from $Mp_{\mathrm{CP}}$ to $1+(Mp_{\mathrm{CP}}-1)\kappa$ without introducing additional interference.
Let $\gamma_1$ denote the SINR of single-satellite transmission, i.e., the SINR in \eqref{eq:spectral_efficiency_analysis} with $\mathrm{PFD}_0 \cdot M \cdot p_{\mathrm{CP}}$ replaced by $\mathrm{PFD}_0$.
In the low-SINR regime of D2H links, $\log_2(1+x)\approx x/\ln 2$, and the proposed strategy outperforms single-satellite transmission when
\begin{equation}
\kappa \;\ge\; \frac{L_{\mathrm{CP}}^{\mathrm{add}}}{(N + L_{\mathrm{CP}})(M p_{\mathrm{CP}} - 1)}.
\label{eq:condition_outperforms}
\end{equation}
Since $Mp_{\mathrm{CP}}$ is typically on the order of tens, \eqref{eq:condition_outperforms} is satisfied unless the phase error is nearly uniformly distributed.
In practice, distributed beamforming prototypes keep the residual phase error within about $15^\circ$~\cite{TWC13_DistBF_Proto}, and the centimeter-level inter-satellite ranging available with commercial off-the-shelf radios corresponds to about $25^\circ$ at 2~GHz~\cite{IFAC25_MultiSatControl}, so that $\kappa$ exceeds 0.76 for $\sigma_\zeta \le 30^\circ$.

Residual Doppler errors are handled in two steps.
The Doppler shift is pre-compensated at the beam level by the network~\cite{38.821}, and the residual within a beam is known from the ephemeris and the UT position; its phase progression over OFDM symbols is compensated in the precoder in the same manner as the delay term in \eqref{eq:analysis_precoder}.
The remaining rotation within a symbol causes ICI with a power of approximately $(\pi\xi)^2/3$ relative to the desired signal~\cite{TCOM95_CFO}, where $\xi$ is the residual carrier frequency offset (CFO) normalized by the subcarrier spacing.
For a satellite antenna array with a beamwidth of a few degrees, the differential Doppler within the beam is on the order of 1~kHz, i.e., $\xi$ below 0.05 for a subcarrier spacing of 30~kHz, which corresponds to an ICI floor below $-21$~dB.
Oscillator errors within the frequency error requirement of $\pm0.1$~ppm~\cite{3gpp381015} give $\xi<0.007$ and an ICI floor below $-38$~dB.
Both floors are far below the SINR of D2H links and affect single- and multi-satellite transmission alike.

\subsection{Practical Considerations}
The proposed strategy is UT-centric.
Each UT determines its attachable satellites by Algorithm~\ref{alg:downlink_sync} from the ephemeris and its own position, both of which are already available in Rel-17 NTN, and reports the association set together with the number of associated satellites $|\mathcal{M}_k|$ required for the power normalization in \eqref{eq:normalized_power} during random access.
Each satellite transmits without exchanging UT-specific information with the other satellites, using its own channel and timing offset to the UT.
The precoder in \eqref{eq:analysis_precoder} and the symbol-index adjustment depend only on the satellite's own geometry and delay, so that a satellite does not need to know which other satellites serve the same UT, and each satellite requires only its own CSI to the UT as in single-satellite transmission.
The association set is updated as the satellites move, and $L_{\mathrm{margin}}$ keeps the association valid between updates.

Joint transmission requires that the data and the modulation and coding scheme (MCS) of a UT be available at all associated satellites.
This is provided by the gateway through the feeder links or inter-satellite links (ISLs)~\cite{VTM21_StarlinkISL}, in the same manner as the central processing unit in cell-free architectures~\cite{OJCOMS22_MultiLEO_narrow, JSAC25_MultiLEO_narrow, TWC17_cellfree}.
The common symbol timing among satellites can be provided by GNSS-based timing~\cite{38.821} or inter-satellite time synchronization~\cite{OJCOMS22_MultiLEO_narrow}, and residual timing errors are absorbed by $L_{\mathrm{margin}}$.
The timing information for each UT is provided by the UT itself during random access.
Coherent combining further requires carrier-phase coherence among the associated satellites, which is assumed in this paper as in~\cite{OJCOMS22_MultiLEO_narrow, TVT25_MultiLEO_wide}.
The sample-level timing offsets addressed in this paper remain even under perfect synchronization among satellites, since they cannot be compensated for multiple UTs simultaneously under the 3GPP architecture.

In terms of latency, the proposed strategy involves the extended symbol duration and the delivery of data to the associated satellites.
Over the air, the symbols from the associated satellites are aligned within the CP window, so the only cost is the symbol duration extended by the additional CP, which is on the order of tens of microseconds per symbol and negligible compared with the propagation delay of several milliseconds.
The alignment is achieved by the symbol-index adjustment at each satellite, which requires the data to be available at the farther satellites earlier.
Since the delays of the feeder links and ISLs described above are known from the ephemeris, the delivery can be scheduled accordingly, and the resulting latency is that of the longest path among the associated satellites.
The additional delay relative to single-satellite transmission is therefore bounded by the spread of the path delays among the visible satellites, and it is a fixed offset that does not accumulate over symbols.
Moreover, the association and the timing report are completed during random access, so no additional round trip is required.
Note that the higher SINR achieved by cooperation reduces the number of retransmissions, each of which costs a round-trip delay over the satellite link.

\section{Simulation Results}\label{sec_6}
This section evaluates the performance of the proposed satellite association strategy for multi-LEO satellite D2H communication.
We consider a scenario with LEO satellites at an altitude of 600~km and a minimum elevation angle of $10^\circ$~\cite{38.811}.
The satellites are uniformly distributed within a region characterized by a central angle of $15.84^\circ$, and the UTs are uniformly distributed within a ground region defined by the same central angle.
Here, the central angle refers to the Earth-centered angular extent of the region in which satellites or UTs can be located.
This scenario is chosen because all satellites can be positioned above the minimum elevation angle of all UTs, thereby allowing each UT to potentially associate with every satellite.

Since the focus is on D2H communication, we adopt the S-band ($2$\;GHz) carrier frequency with a 30\;MHz bandwidth scenario, as specified in 3GPP TR 38.821~\cite{38.821}. 
Furthermore, in accordance with the regulations of the ITU, the received power on the ground is subject to a limit on the PFD~\cite{ITU}.
To comply with the PFD regulation, the average received power at the UT is normalized to remain identical across single- and multi-satellite transmission scenarios. 
Specifically, if the received power is $P$ with a single satellite, then under joint transmission with $M$ satellites, each satellite contributes $P/M$, such that the aggregate non-coherent received power remains $P$. 
It should be emphasized that coherent combining among satellites may increase the effective received power at a UT. 
However, this does not violate the PFD regulation, since the PFD limit is determined by the transmit power and beam pattern of each satellite rather than by the constructive or destructive superposition of signals at the receiver~\cite{ITU}.
The PFD limit, along with other simulation parameters, is summarized in Table~\ref{tab:chap3_simulation_parameter}.

The results are obtained by Monte Carlo simulation over independent realizations of the satellite and UT positions, each containing all $M$ satellites and $K$ UTs, with the number of realizations given in Table~\ref{tab:chap3_simulation_parameter}.
In each realization, the sample offsets are computed by \eqref{eq:sample_offset} with the sampling period in Table~\ref{tab:chap3_simulation_parameter}, the downlink synchronization point of each UT is determined by \eqref{eq:downlink_synchronization}, or drawn uniformly at random under random synchronization, and the association set is formed according to each scheme.
The throughput of each UT is then computed with the channel in \eqref{eq:simplified_channel} and the precoder in \eqref{eq:analysis_precoder}, using the SINR in \eqref{eq:SINR_kn} and the throughput in \eqref{eq:Rk_DL} for the baselines, and the power terms in \eqref{eq:modified_power_terms} and the throughput in \eqref{eq:modified_total_spectral_efficiency} for the proposed strategy.
The average throughput is taken over all UTs and realizations, and the empirical cumulative distribution function (ECDF) is obtained from the throughput of all UTs in all realizations.

\begin{figure}[!t]
    \centering
    \includegraphics[width=1\columnwidth]{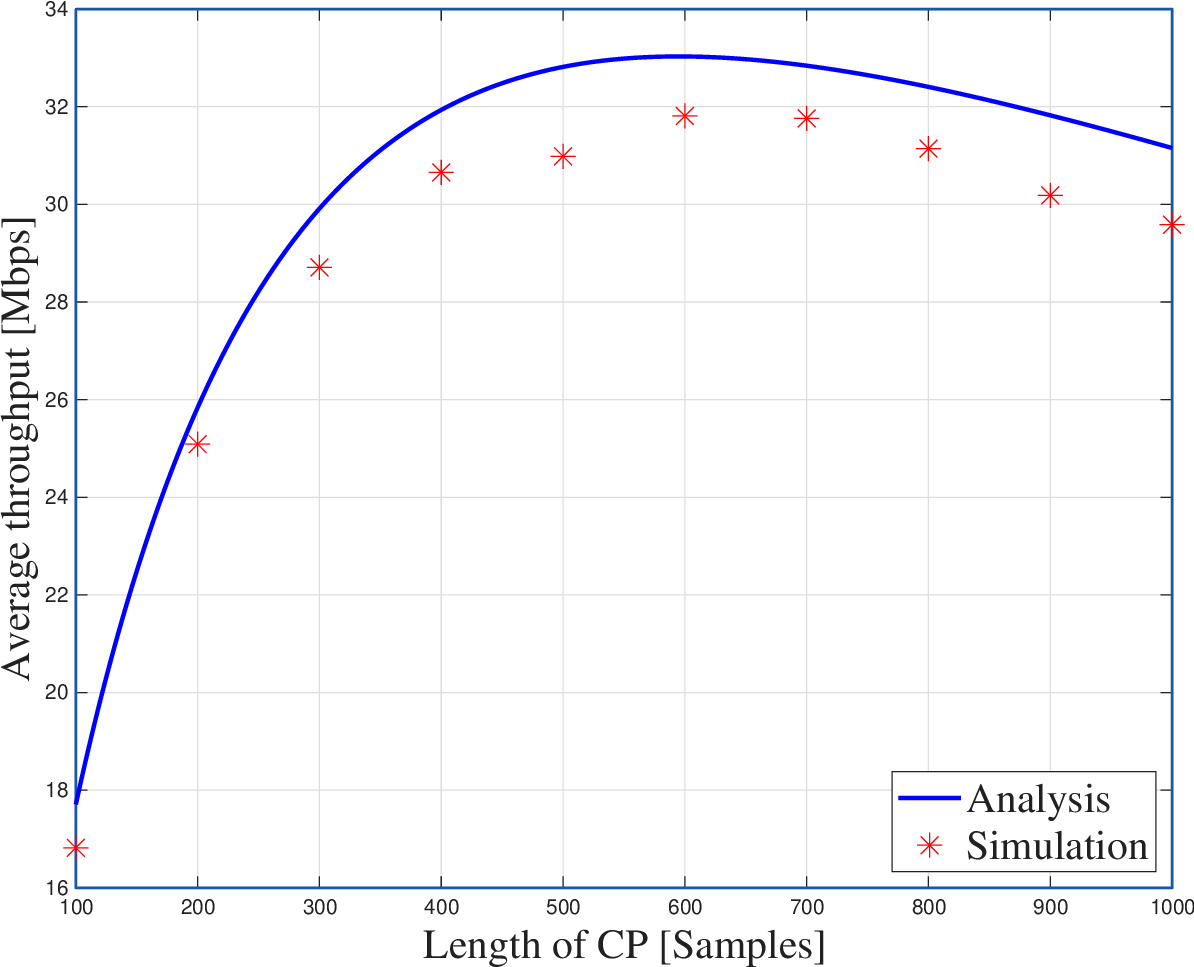}
    \caption{Average throughput performance of the proposed strategy as a function of the additional CP length.}
    \label{fig:se_CP}
\end{figure}

To focus on the performance impact of satellite association and cooperation, we assume simple transmit/receive beamformers based on MRT and maximum ratio combining (MRC). 
For evaluation, the following schemes are considered to demonstrate the effectiveness of the proposed satellite association strategy.
\begin{itemize}
    \item \textbf{Single-satellite system:} Each UT is connected only to the nearest satellite and receives service exclusively from it. 
    From the perspective of an individual satellite, this setup corresponds to a multi-user MIMO (MU-MIMO) system serving multiple UTs simultaneously~\cite{JSAC20_singleSAT, TCOM22_LOS}.
    \item \textbf{Fully connected multi-satellite system:} Each UT is connected to all satellites that satisfy the minimum elevation angle constraint. 
    Similar to a cell-free terrestrial system, each UT receives joint transmission from all accessible satellites without any cell boundaries~\cite{TWC17_cellfree}.
    \item \textbf{Proposed multi-satellite system:} Each UT allocates an additional CP within the OFDM symbol and associates only with satellites whose timing offsets fall within this extended CP. 
    Compared to the single-satellite system, more satellites can be associated, while fewer satellites are connected than in the fully connected case. 
    Unless otherwise specified, the additional CP length $(L_{\mathrm{CP}}^{\mathrm{add}})$ is set to 600, and downlink synchronization is optimized to maximize the number of associated satellites.
    This value is selected based on Fig.~\ref{fig:se_CP}, where the throughput is maximized around $L_{\mathrm{CP}}^{\mathrm{add}} = 600$ under the considered setting.
    It should be noted that the additional CP reduces the pre-log factor in \eqref{eq:modified_total_spectral_efficiency} to $N/(N + L_{\mathrm{CP}} + L_{\mathrm{CP}}^{\mathrm{add}}) \approx 0.61$, and that all throughput results in this section include this loss.
\end{itemize}

\begin{table}[!t]
\centering
\caption{Simulation parameter setting.}
\label{tab:chap3_simulation_parameter}
\renewcommand{\arraystretch}{1.2}
\resizebox{1\columnwidth}{!}{
\begin{tabular}{l l}
\toprule
\textbf{Simulation Parameters} & \textbf{Value} \\
\midrule
Satellite altitude & 600\;km \\
Minimum elevation angle ($\epsilon_\mathrm{min}$) & $10^\circ$ \\
SAT maximum central angle & $15.84^\circ$ \\
UT maximum central angle & $15.84^\circ$ \\
Center frequency ($f_{\mathrm{c}}$) & 2\;GHz \\
Bandwidth & 30\;MHz \\
FFT size ($N$) & 1,024 \\
Sampling period ($T_{\mathrm{s}}$) & 33.3\;ns \\
Satellite antenna array & $32 \times 32$ \\
UT antenna array & $1 \times 1$ \\
Antenna element spacing ($d_{\mathrm{s}}$) & $\lambda/2$ \\
CP length ($L_{\mathrm{CP}}$) & 64 \\
Additional CP length ($L_{\mathrm{CP}}^{\mathrm{add}}$) & 600 \\
PFD limit~\cite{ITU} & $-144~\text{dBW}/\text{m}^2/4\;\text{kHz}$ \\
Noise power per subcarrier ($\sigma_w^2$)~\cite{38.821} & $-152.24~\text{dBW}$ \\
Total number of satellites & 2:2:8, 10:10:100 \\
Total number of UTs & 10:10:100 \\
Phase error standard deviation ($\sigma_\zeta$) & $0^\circ$:$15^\circ$:$180^\circ$ \\
Number of realizations & $10^4$ \\
\bottomrule
\end{tabular}}
\end{table}

\begin{figure}[!t]
    \centering
    \includegraphics[width=1\columnwidth]{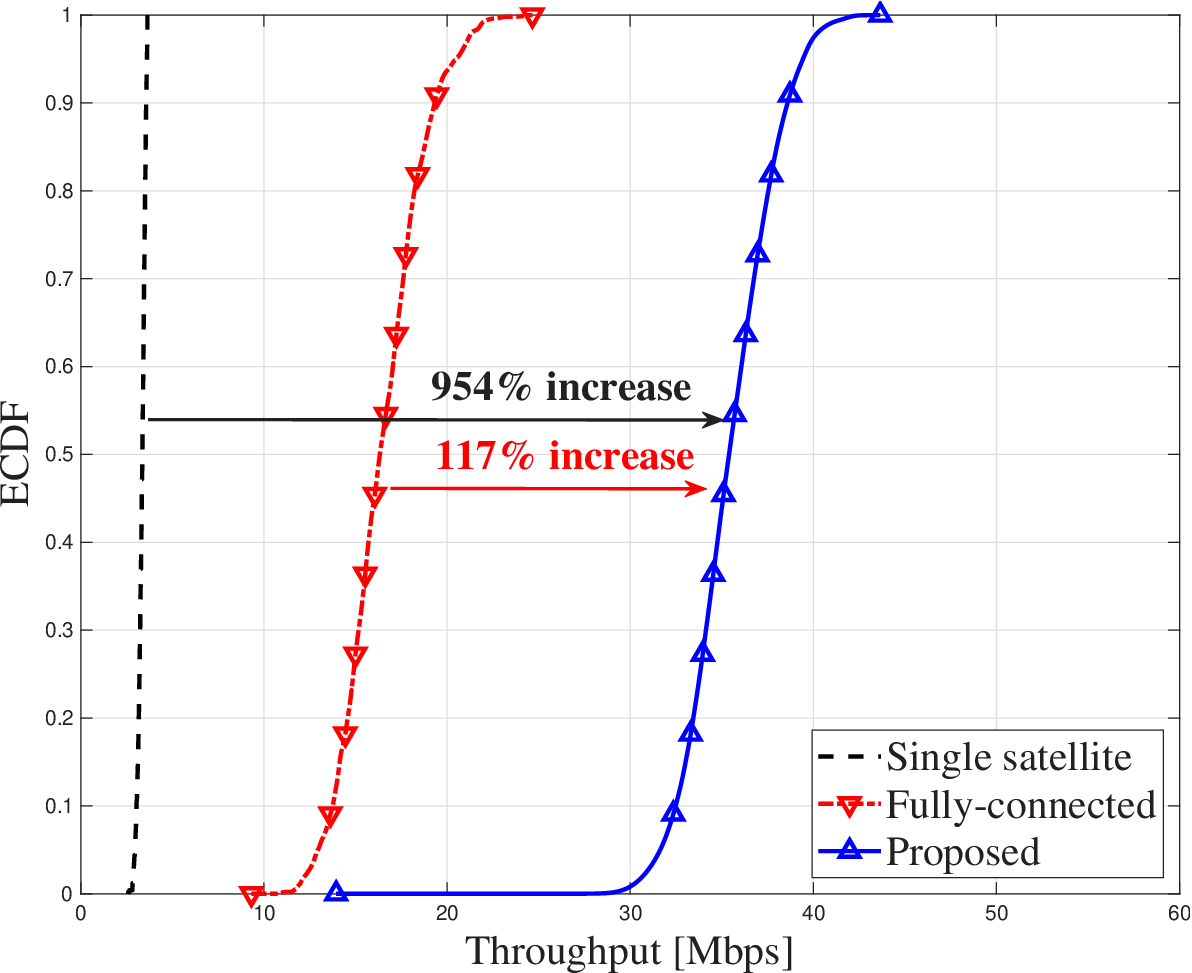}
    \caption{ECDF of the throughput per UT with 100 satellites and 100 UTs.}
    \label{fig:spectral_efficiency_ECDF}
\end{figure}

Fig.~\ref{fig:spectral_efficiency_ECDF} illustrates the ECDF of the UT throughput when 100 satellites and 100 UTs are deployed.
Based on the median values, the throughput performance ranks in the order of the proposed multi-satellite system, the fully connected multi-satellite system, and the single-satellite system.
Interestingly, in the fully connected case, although the desired signal power increases with the number of cooperating satellites, the effects of self-ISI and self-ICI become dominant, preventing proportional performance improvement.
In contrast, the proposed multi-satellite strategy mitigates the dominant ICI and ISI caused by the desired signal itself and thereby achieves a throughput improvement of approximately $954\%$ compared to the single-satellite case.
These results demonstrate that forming a cooperative network through the proposed strategy enables multi-LEO satellite joint transmission to provide sufficient power gain.

\begin{figure}[!t]
    \centering
    \includegraphics[width=1\columnwidth]{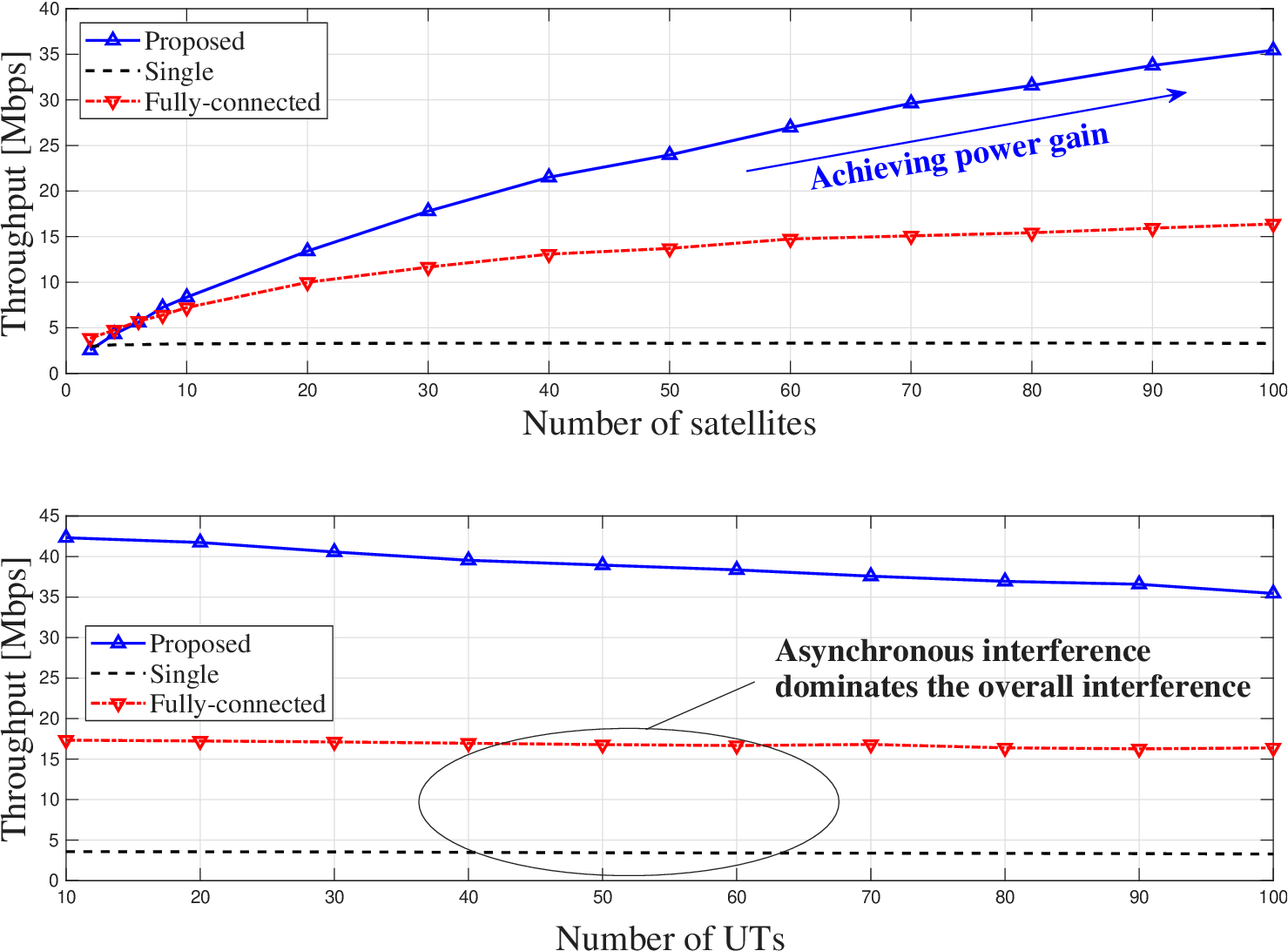}
    \caption{Average throughput performance of different systems. The upper plot shows the throughput versus the number of satellites with 100 UTs, while the lower plot shows the throughput versus the number of UTs with 100 satellites.}
    \label{fig:spectral_efficiency_SAT_UT}
\end{figure}

Fig.~\ref{fig:spectral_efficiency_SAT_UT} presents the average throughput as a function of the number of satellites and the number of UTs.
For the proposed strategy, throughput increases with the number of satellites, indicating that the power gain from joint transmission is effectively exploited.
In contrast, the single-satellite case is limited by serving only one satellite, and the fully connected multi-satellite case fails to capture power gains due to the significant interference effects.
Moreover, even under a practical constraint where the number of attachable satellites per UT is limited to around ten, the proposed strategy achieves an average throughput that is 15.8\% higher than the fully connected baseline and 157.8\% higher than the single-satellite case.

The upper plot of Fig.~\ref{fig:spectral_efficiency_SAT_UT} also covers the sparse regime with two to eight satellites.
As observed in the figure, the proposed strategy falls below the fully connected case for up to about six satellites and below the single-satellite case with two satellites.
With so few satellites, the asynchronous self-interference removed by the association is small, whereas the additional CP costs the same time resources, so connecting to all visible satellites without the additional CP is preferable.
The advantage of the proposed strategy appears as the number of satellites grows, since the excluded satellites then contribute more self-interference than power gain, and it increases with the satellite density.
Since the UT knows the number of visible satellites from the ephemeris, it can fall back to the fully connected or single-satellite operation in such sparse deployments.

When examining throughput as a function of the number of UTs, all cases experience degradation as the number of UTs increases, owing to stronger MUI as well as ICI and ISI caused by other UTs. 
In the fully connected multi-satellite case, however, the asynchronous interference from the desired signals is already dominant, so the additional impact of MUI growth is marginal, resulting in only minor changes in spectral efficiency with increasing UTs.
This also indicates that the interference in the fully connected case is self-generated and cannot be removed by interference-suppressing precoders.

Fig.~\ref{fig:se_CP} illustrates the average throughput of the proposed strategy as a function of the CP length. 
The blue curve represents the upper bound derived from \eqref{eq:spectral_efficiency_analysis}, while the red star markers denote the simulation results.
Both the analytical and simulation outcomes are obtained under random downlink synchronization.
As observed in the figure, the analytical upper bound closely matches the simulation trend.
This agreement validates the analytical model and enables a reliable interpretation of the throughput behavior for different CP lengths.
For relatively short CP lengths, the system operates in a low-SINR regime, where the throughput is highly sensitive to power gain; thus, the throughput improves as the CP length increases. 
However, when the CP length exceeds approximately 600 samples, the reduction in time-resource efficiency caused by the pre-log term becomes more dominant than the SINR enhancement obtained through additional power gain, resulting in a decrease in throughput. 
Therefore, the CP length should be carefully determined by properly accounting for this trade-off in conjunction with the satellite and UT distribution and the FFT size.
Specifically, for the parameters in Table~\ref{tab:chap3_simulation_parameter}, $\gamma_M$ in \eqref{eq:optimal_CP} is about 9.4~dB, which gives $L_{\mathrm{CP}}^{\mathrm{add}\ast} \approx 550$ samples and agrees with the figure. The throughput is within 1\% of its maximum between about 450 and 700 samples, so a fixed CP length is robust to moderate changes in the deployment.

\begin{figure}[!t]
    \centering
    \includegraphics[width=1\columnwidth]{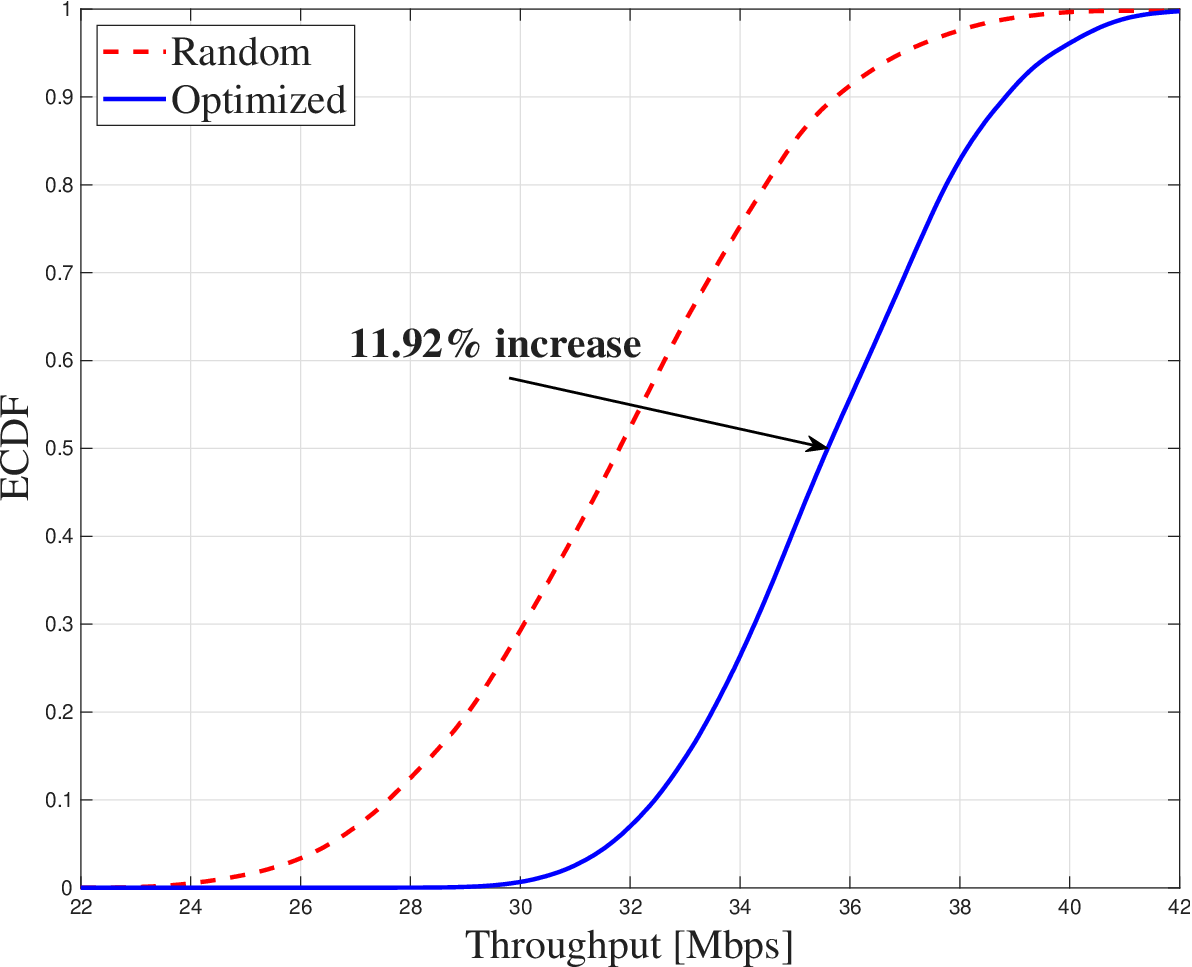}
    \caption{Comparison of throughput with optimized versus random downlink synchronization.}
    \label{fig:spectral_efficiency_opt}
\end{figure}

Fig.~\ref{fig:spectral_efficiency_opt} compares the throughput achieved with optimized downlink synchronization and with random downlink synchronization, as introduced in Section~\ref{sec_5}.C.
By performing proper downlink synchronization, the number of associated satellites can be maximized, thereby achieving $11.92\%$ higher throughput at the median.
Without proper synchronization, the UT can only connect to fewer satellites than the maximum available, resulting in insufficient power gain from multiple satellites.
Therefore, optimized downlink synchronization is essential to fully exploit the power gain of the desired signal from multiple satellites.

\begin{figure}[!t]
    \centering
    \includegraphics[width=1\columnwidth]{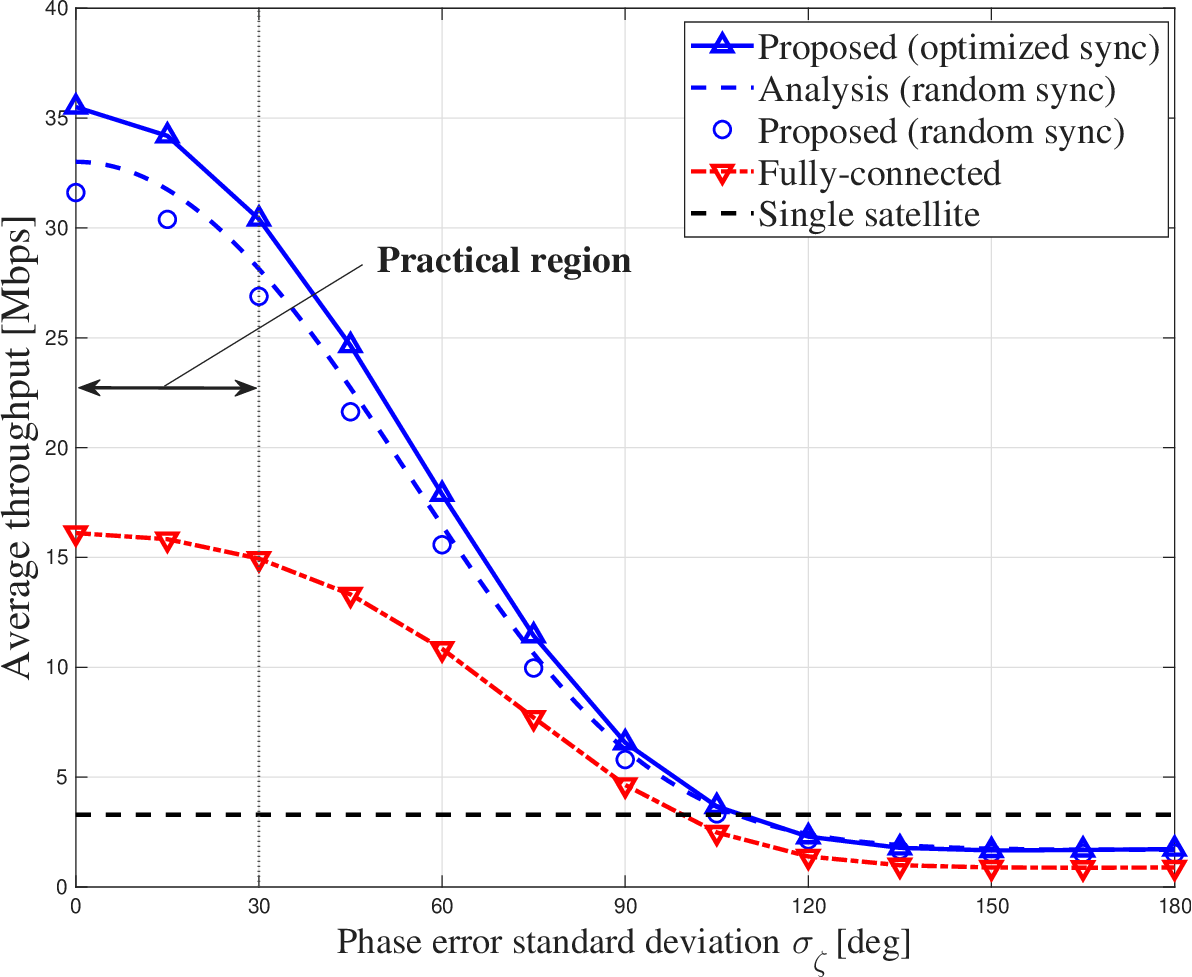}
    \caption{Average throughput performance of different systems as a function of the phase error standard deviation $\sigma_\zeta$ with 100 satellites and 100 UTs. The shaded area marks the practical region $\sigma_\zeta \le 30^\circ$.}
    \label{fig:phase_error}
\end{figure}

Fig.~\ref{fig:phase_error} illustrates the average throughput as a function of the standard deviation of the phase error $\sigma_\zeta$ when 100 satellites and 100 UTs are deployed.
The dashed curve represents the analytical result obtained from \eqref{eq:spectral_efficiency_analysis} with \eqref{eq:desired_power_error} under random synchronization, the circle markers denote the corresponding simulation results of the proposed strategy, in which the precoder of each satellite is rotated by an i.i.d. Gaussian phase error, and the solid curve denotes the simulation results when synchronization is performed according to \eqref{eq:downlink_synchronization}.
As observed in the figure, the simulation results closely match the analytical curve over the entire range, which validates the phase error model in \eqref{eq:desired_power_error}.
The single-satellite case is not affected by the phase error, since a phase rotation common to the entire signal does not change the received power.
In contrast, the throughput of the proposed strategy decreases gradually with $\sigma_\zeta$, as the coherent power gain from joint transmission is partially lost, and it falls below the single-satellite case between $105^\circ$ and $120^\circ$, which agrees with the condition in \eqref{eq:condition_outperforms}.
The fully connected case decreases in the same manner and remains below the proposed strategy over the entire range, since it suffers from both the phase error and the asynchronous interference.
The shaded area marks the practical region $\sigma_\zeta \le 30^\circ$, which covers the residual phase errors of distributed beamforming in practice.
Within this region, the proposed strategy retains 96\% and 86\% of its perfect-CSI throughput at $\sigma_\zeta = 15^\circ$ and $30^\circ$, which demonstrates that the power gain of the proposed strategy is robust to imperfect CSI.

The results above clarify the trade-off between single-satellite and multi-satellite cooperative transmission.
Single-satellite transmission requires neither the additional CP nor the data distribution and carrier-phase coherence among satellites, but its received power is limited by the PFD constraint of a single satellite.
Multi-satellite cooperation increases the desired power in proportion to the number of attachable satellites, at the cost of the time-resource efficiency reduced by the additional CP, the coordination among satellites, and the sensitivity to the phase error among them.
Consequently, cooperation is beneficial whenever more than one satellite is visible, and the proposed strategy becomes the preferred form of cooperation beyond about six visible satellites in the considered setting, with a gain that grows with the satellite density.

\section{Conclusion}\label{sec_7}
This paper addressed multi-LEO cooperative D2H communication by focusing on the fundamental impact of inter-satellite timing mismatch under the 3GPP MIMO-OFDM structure. 
We showed that satellite-dependent propagation delays inevitably generate ICI and ISI, where the desired signal itself can become a dominant interference source. 
To mitigate this issue without requiring additional timing synchronization among satellites, we proposed a timing-aware satellite association strategy that introduces an additional CP and restricts associations to satellites whose delay offsets remain within the extended CP margin. 
Simulation results confirmed that the proposed strategy significantly improves spectral efficiency compared to single-satellite transmission and fully connected multi-satellite baselines.

Future work includes advanced beamforming and interference suppression to further reduce residual MUI and asynchronous ICI/ISI, mobility-aware association and scheduling to cope with rapidly varying LEO geometry, and adaptive additional-CP design to balance connectivity gains and CP overhead under diverse deployment conditions.

\bibliographystyle{IEEEtran}
\bibliography{reference}





\end{document}